# False-Name Manipulations in Weighted Voting Games

**Haris Aziz**                                                    AZIZ@IN.TUM.DE
*Institut für Informatik*
*Technische Universität München, Germany*

**Yoram Bachrach**                                          YORAMBAC@GMAIL.COM
*Microsoft Research*
*Cambridge, UK*

**Edith Elkind**                                              EELKIND@NTU.EDU.SG
*School of Physical and Mathematical Sciences*
*Nanyang Technological University, Singapore*

**Mike Paterson**                                        MSP@DCS.WARWICK.AC.UK
*Department of Computer Science*
*University of Warwick, UK*

## Abstract

Weighted voting is a classic model of cooperation among agents in decision-making domains. In such games, each player has a weight, and a coalition of players wins the game if its total weight meets or exceeds a given quota. A player's power in such games is usually not directly proportional to his weight, and is measured by a power index, the most prominent among which are the Shapley–Shubik index and the Banzhaf index.

In this paper, we investigate by how much a player can change his power, as measured by the Shapley–Shubik index or the Banzhaf index, by means of a *false-name manipulation*, i.e., splitting his weight among two or more identities. For both indices, we provide upper and lower bounds on the effect of weight-splitting. We then show that checking whether a beneficial split exists is NP-hard, and discuss efficient algorithms for restricted cases of this problem, as well as randomized algorithms for the general case. We also provide an experimental evaluation of these algorithms.

Finally, we examine related forms of manipulative behavior, such as annexation, where a player subsumes other players, or merging, where several players unite into one. We characterize the computational complexity of such manipulations and provide limits on their effects. For the Banzhaf index, we describe a new paradox, which we term the Annexation Non-monotonicity Paradox.

## 1. Introduction

Collaboration and cooperative decision-making are important issues in many types of interactions among self-interested agents (Ephrati & Rosenschein, 1997). In many situations, agents must take a joint decision leading to a certain outcome, which may have a different impact on each of the agents. A standard and well-studied way of doing so is by means of voting, and in recent years, there has been a lot of research on applications of voting to multiagent systems as well as on computational aspects of various voting procedures (see Faliszewski & Procaccia, 2010). One of the key issues in this domain is how to measure the *power* of each voter, i.e., his impact on the final outcome. In particular, this question becomes important when the agents have to decide how to distribute the payoffs resulting from their joint action: a natural approach would be to pay each agent according to his contribution, i.e., his voting power.

This issue is traditionally studied within the framework of *weighted voting games (WVGs)* (Taylor & Zwicker, 1999), which provide a model of decision-making in many political and legislative bodies (Leech, 2002; Laruelle & Widgren, 1998; Algaba, Bilbao, & Fernández, 2007), and have also been investigated in the context of multiagent systems (Elkind, Goldberg, Goldberg, & Wooldridge, 2008b, 2007). In such a game, each of the agents has a weight, and a coalition of agents wins the game if the sum of the weights of its participants meets or exceeds a certain quota. There are numerous examples of multiagent systems that can be captured by weighted voting games. For example, the agents' weights may correspond to the amount of resources (time, money, or battery power) that they contribute, and the quota may indicate the amount of resources needed to complete a given task. Alternatively, the weight may be an indicator of an agent's experience or seniority, and a voting procedure may be designed to take into account these characteristics.

Clearly, having a larger weight makes it easier for a player to affect the outcome. However, the player's power is not always proportional to his weight. For example, if the quota is so high that the only winning coalition is the one that includes all players, intuitively, all players have equal power, irrespective of their weight. This idea is formalized using the concept of a *power index*, which is a systematic way of measuring a player's influence in a weighted voting game. There are several ways to define power indices. One of the most popular approaches relies on the fact that weighted voting games form a subclass of *coalitional games*, and therefore one can use the terminology and solution concepts that have been developed in the context of general coalitional games. In particular, an important notion in coalitional games is that of the Shapley value (Shapley, 1953), which is a classic method of distributing the gains of the grand coalition in general coalitional games. The Shapley value has a natural interpretation in the context of weighted voting, where it is known as the Shapley–Shubik power index (Shapley & Shubik, 1954). Another well-known power index, which has been introduced specifically in the context of weighted voting games, is the Banzhaf index (Banzhaf, 1965). While several other power indices have been proposed (e.g., see Johnston, 1978; Deegan & Packel, 1978; Holler & Packel, 1983), the Shapley–Shubik power index and the Banzhaf power index are usually viewed as the two standard approaches to measuring the players' power in weighted voting games, and have been widely studied from both normative and computational perspective.

As suggested above, power indices measure the players' power and can be used to determine their payoffs. However, to be applicable in real-world scenarios, this approach to payoff division has to be resistant to dishonest behavior, or manipulation, by the participating players. In this paper we study the effects of a particular form of manipulation in weighted voting games, namely, false-name voting. Under this manipulation, a player splits his weight between himself and a "fake" agent who enters the game. Such manipulations are virtually impossible to detect in open anonymous environments such as the internet. They can also occur in legislative bodies, where political parties vote on bills. In such bodies, elections are held every several years to determine the weight each party has when voting on a bill. Before such elections are held, a party may split up into two smaller parties. It is likely that the supporters of that party would somehow split between the two new parties, so the total weight of the new parties will be equal to that of the original party. By choosing a suitable platform, the original party can decide how the weight would be split between the two new parties.

While a weight-splitting manipulation does not change the total weight of all identities of the cheating agent, his power (as measured by the Shapley–Shubik power index or the Banzhaf index) may change. Therefore, this behavior presents a challenge to the designers of multiagent systems that rely on weighted voting. The main goal of this paper is to measure the effects of false-name voting and analyze its computational feasibility. We also examine the related scenarios of players merging in order to increase their joint power, or one player annexing another one.

Our main results are as follows:

- We precisely quantify the worst-case effect of false-name voting on agents' payoffs. Namely, we show that in an $n$-player game splitting into two false identities can increase an agent's payoff by at most a factor of 2 both for the Shapley–Shubik index and the Banzhaf index. Moreover, this bound is asymptotically tight. On the other hand, we show that false-name manipulation can decrease an agent's payoff by at most a factor of $\Theta(n)$ for both indices.

- We demonstrate that finding a successful manipulation is not a trivial task by proving that for both indices it is NP-hard to verify if a beneficial split exists. However, we show that if all weights are polynomially bounded, the problem can be solved in polynomial time, and discuss efficient randomized algorithms for this problem.

- We present similar NP-hardness results for the case of players merging into a single new player. Interestingly, in the case of a player annexing one or more players, there is a contrast between the Shapley–Shubik index and the Banzhaf index. Whereas for the Shapley–Shubik index, annexing is always beneficial, checking whether annexing is beneficial in the case of the Banzhaf index is NP-hard. However it is beneficial if a player annexes a player with a bigger weight. We also present a new paradox called *The Annexation Non-monotonicity Paradox*, which shows that annexing a "small" player can be more useful than annexing a "big" player.

- We complement our theoretical results by experiments which indicate the expected fractions of positive and negative false-name manipulations in weighted voting games with randomly selected weights.

## 1.1 Related Work

Weighted voting games date back at least to John von Neumann and Oskar Morgenstern, who developed their theory in their monumental book *Theory of Games and Economic Behavior* (von Neumann & Morgenstern, 1944). Subsequently, WVGs have been analyzed extensively in the game theory literature (see, for instance, Taylor & Zwicker, 1999).

In his seminal paper, Shapley (1953) considered coalitional games and the question of fair allocation of the utility gained by the grand coalition. The solution concept introduced in this paper became known as the *Shapley value* of the game. The subsequent paper (Shapley & Shubik, 1954) studies the Shapley value in the context of simple coalitional games, where it is usually referred to as the *Shapley–Shubik power index*. The Banzhaf power index was originally introduced by Banzhaf (1965); a somewhat different definition was later proposed by Dubey and Shapley (1979). In this paper, we make use of Banzhaf's original definition, as it is more appropriate in the context of payoff division.

Both of these power indices have been well studied. Straffin (1977) shows that each index reflects certain conditions in a voting body. Laruelle (1999) describes certain axioms that characterize these two indices, as well as several others. These indices were used to analyze the voting structures of the European Union Council of Ministers and the IMF (Machover & Felsenthal, 2001; Leech, 2002).

The applicability of the power indices to measuring political power in various domains has raised the question of finding tractable ways to compute them. However, this problem appears to be computationally hard. Indeed, the naive algorithm for calculating the Shapley value (or the Shapley–Shubik power index) considers all permutations of the players and hence runs in exponential time. Moreover, Papadimitriou and Yannakakis (1994) show that computing the Shapley value in weighted voting games is #P-complete. This result is extended by Matsui and Matsui (2001), who show that calculating the Banzhaf index in weighted voting games is also NP-hard. Furthermore, Faliszewski and Hemaspaandra (2009) show that comparing the player's power in two different weighted voting games is PP-complete for both indices.

Despite these hardness results, several papers show how to compute these power indices in some *restricted domains*, or discuss ways to *approximate* them. These include a generating functions approach (Mann & Shapley, 1962), which trades required storage for running time, Owen's multilinear extension (MLE) approach (Owen, 1975) and Monte Carlo simulation approaches (Mann & Shapley, 1960; Fatima, Wooldridge, & Jennings, 2007; Bachrach, Markakis, Resnick, Procaccia, Rosenschein, & Saberi, 2010). Matsui and Matsui (2000) provide a good survey of algorithms for calculating power indices in weighted voting games. Many of these approaches work well in practice, which justifies the use of these indices as payoff distribution schemes in multiagent domains.

As a useful and succinct model for coalitional voting games, WVGs have attracted a lot of interest from the multiagent community. A number of papers have considered the problem of designing WVGs with desirable properties (Aziz, Paterson, & Leech, 2007; Fatima, Wooldridge, & Jennings, 2008; de Keijzer, Klos, & Zhang, 2010). Simple games that can be obtained by combining multiple weighted voting games have been examined by Elkind et al. (2008b) and Faliszewski, Elkind, and Wooldridge (2009). Another well-studied topic is computing various stability-related solution concepts in WVGs and their extensions (Elkind et al., 2007; Elkind & Pasechnik, 2009; Elkind, Chalkiadakis, & Jennings, 2008a).

False-name manipulations in open anonymous environments have been examined in different domains such as auctions (Yokoo, 2007; Iwasaki, Kempe, Saito, Salek, & Yokoo, 2007) and coalitional games (Yokoo, Conitzer, Sandholm, Ohta, & Iwasaki, 2005; Ohta, Iwasaki, Yokoo, Maruono, Conitzer, & Sandholm, 2006; Ohta, Conitzer, Satoh, Iwasaki, & Yokoo, 2008). In the latter domain, the characteristic function by itself does not provide enough information to analyze false-name manipulation. To deal with this issue, Yokoo et al. (2005) introduced a framework where each player has a subset of skills, and the characteristic function assigns values to the subset of skills. Our model can be seen as a special case of this framework; however, due to special properties of weighted voting games we are able to obtain much stronger results than in the general case.

The phenomenon considered in the paper has been studied by political scientists and economists under the name of "the paradox of size" (Shapley, 1973; Brams, 1975; Felsenthal & Machover, 1998); however, neither its quantitative nor its computational aspects have been considered. Felsenthal and Machover also discuss a number of other paradoxes in weighted voting games; Laruelle and Valenciano (2005) give an overview of more recent work on paradoxes of weighted voting. Occurrences of these paradoxes in voting bodies are considered by Kilgour and Levesque (1984),van Deemen and Rusinowska (2003), and Leech and Leech (2005). Another form of manipulation in WVGs has been recently studied by Zuckerman, Faliszewski, Bachrach, and Elkind (2008), who analyze how the center might change the players' power by modifying the quota even if the weights are fixed.

### 1.2 Follow-up Work

Many results that appear in this paper have been previously presented at the AAMAS conference (Bachrach & Elkind, 2008; Aziz & Paterson, 2009). Inspired by this work, Lasisi and Allan (2010) have recently undertaken an experimental analysis of false-name manipulations in weighted voting games. They have also considered less popular power indices, such as the Deegan–Packel index. In another follow-up paper, Rey and Rothe (2010) investigate false-name manipulations in weighted voting games with respect to the probabilistic Banzhaf index, i.e., the one suggested by Dubey and Shapley (1979). Although the probabilistic Banzhaf index is more useful for measuring the actual probability of influencing a decision, it does not fit the framework of using power indices to share resources or power, because the probabilistic Banzhaf index is not normalized.

## 2. Preliminaries and Notation

We start by introducing the notions that will be used throughout this paper.

### 2.1 Coalitional Games

A *coalitional game* $G = (N, v)$ is given by a set of *players* $N = \{1, \ldots, n\}$, and a *characteristic function* $v : 2^N \to \mathbb{R}$, which maps any subset, or *coalition*, of players to a real value. This value is the total utility these players can guarantee to themselves when working together.

A coalitional game $G = (N, v)$ is called *monotone* if $v(S) \leq v(T)$ for any $S \subseteq T$. Further, $G$ is called *simple* if it is monotone and $v$ can only take values 0 and 1, i.e., $v : 2^N \to \{0, 1\}$. In such games, we say that a coalition $S \subseteq N$ *wins* if $v(S) = 1$, and *loses* if $v(S) = 0$. A player $i$ is *critical*, or *pivotal*, for a coalition $S$ if adding this player to $S$ turns it from a losing coalition into a winning coalition: $v(S) = 0$, $v(S \cup \{i\}) = 1$. A player $i$ is a *veto* player if he is necessary for forming a winning coalition, i.e., $v(S) = 0$ for any $S \subseteq N \setminus \{i\}$ (for monotone games, this is equivalent to requiring $v(N \setminus \{i\}) = 0$).

### 2.2 Weighted Voting Games

A *weighted voting game* $G$ is a simple game that is described by a vector of players' *weights* $\mathbf{w} = (w_1, \ldots, w_n) \in (\mathbb{R}^+)^n$ and a *quota* $q \in \mathbb{R}^+$. We write $G = [q; w_1, \ldots, w_n]$, or $G = [q; \mathbf{w}]$. In these games, a coalition is winning if its total weight meets or exceeds the quota. Formally, for any $S \subseteq N$ we have $v(S) = 1$ if $\sum_{i \in S} w_i \geq q$ and $v(S) = 0$ otherwise. We will often write $w(S)$ to denote the total weight of a coalition $S$, i.e., $w(S) = \sum_{i \in S} w_i$. Also, we set $w_{\max} = \max_{i=1,\dots,n} w_i$. We will make the standard assumption that $w(N) \geq q$, i.e., the grand coalition is winning. Note that if $q = w(N)$, then any player $i \in N$ is a veto player.

## 2.3 Power Indices

For each player, both her Shapley–Shubik index and her Banzhaf index are determined by this player's expected marginal contribution to all possible coalitions; however, the two indices make use of different probabilistic models.

The Shapley–Shubik index is a specialization of Shapley value—a classic solution concept for coalitional games—to simple games. In more detail, let $\Pi_n$ be the set of all possible permutations (orderings) of $n$ players. Each $\pi \in \Pi_n$ is a one-to-one mapping from $\{1, \dots, n\}$ to $\{1, \dots, n\}$. Denote by $S_\pi(i)$ the set of predecessors of player $i$ in $\pi$, i.e., $S_\pi(i) = \{j \mid \pi(j) < \pi(i)\}$. The Shapley value of the $i$-th player in a game $G = (N, v)$ is denoted by $\varphi_i(G)$ and is given by the following expression:

$$\varphi_i(G) = \frac{1}{n!} \sum_{\pi \in \Pi_n} [v(S_\pi(i) \cup \{i\}) - v(S_\pi(i))]. \tag{1}$$

We will occasionally abuse notation and say that a player $i$ is pivotal for a permutation $\pi$ if it is pivotal for the coalition $S_\pi(i)$.

The Shapley–Shubik power index is simply the Shapley value in a simple coalitional game (and therefore in the rest of the paper we will use these terms interchangeably). In such games the value of each coalition is either 0 or 1, so formula (1) simply counts the fraction of all orderings of the players in which player $i$ is critical for the coalition formed by his predecessors. The Shapley–Shubik power index thus reflects the assumption that when forming a coalition, any *ordering* of the players entering the coalition has an equal probability of occurring, and expresses the probability that player $i$ is critical.

In contrast, the Banzhaf index computes the probability that the player is critical under the assumption that all coalitions of the players are equally likely. Formally, given a game $G = (N, v)$, for each $i \in N$ we denote by $\eta_i(G)$ the number of coalitions for which $i$ is critical in a game $G$. The *Banzhaf index* of a player $i$ in a WVG $G = (N, v)$ is

$$\beta_i(G) = \frac{\eta_i(G)}{\sum_{j \in N} \eta_j(G)}.$$

While there exist several other approaches to determining the players' influence in a game, the Shapley–Shubik index and the Banzhaf index have many useful properties that make them very convenient to work with. We will make use of three of these properties, namely, the *normalization* property, the *symmetry* property, and the *dummy player* property. The normalization property simply states that the sum of Shapley–Shubik indices (or Banzhaf indices) of all players is equal to 1. The symmetry property says that two players $i, j$ that make the same contribution to any coalition, i.e., such that $v(S \cup \{i\}) = v(S \cup \{j\})$ for any $S \subseteq N \setminus \{i, j\}$, have equal values of the index. The dummy player property claims that for a dummy player both indices equal 0, where a player $i$ is called a *dummy* if he contributes nothing to any coalition, i.e., for any $S \subseteq N$ we have $v(S \cup \{i\}) = v(S)$. It is easy to verify from the definitions that both the Shapley–Shubik index and the Banzhaf index have these properties.

## 3. Weight-Splitting: Examples

In real-world situations modeled by weighted voting games, players may be able to split, dividing their resources (weight) arbitrarily among the new identities. The payoff would then be distributed among the agents according to their power in the resulting game. Intuitively, the total payment obtained by the new identities should be equal to the payoff of the original player before the split. However, we will now demonstrate that this is not the case if the payoff is distributed according to either the Shapley–Shubik index or the Banzhaf index.

We first show that players can use weight-splitting to increase their power.

**Example 1. [Advantageous splitting]** Consider the WVG $[6; 2, 2, 2]$. By symmetry, each player has a Banzhaf index of $1/3$, and a Shapley–Shubik index of $1/3$. If the last player splits up into two players, the new game is $[6; 2, 2, 1, 1]$. In this game, just as in the original game, the only winning coalition is the grand coalition, and hence all players are equally powerful. Thus, the split-up players have a Banzhaf index of $1/4$ each, as well as a Shapley–Shubik index of $1/4$ each, i.e., weight-splitting increases the manipulator's power by a factor of $(2 \cdot 1/4)/(1/3) = 3/2$ according to both indices.

However, weight-splitting may also be harmful.

**Example 2. [Disadvantageous splitting]** Consider the WVG $[5; 2, 2, 2]$. Again, by symmetry, each player has a Banzhaf index of $1/3$, and a Shapley–Shubik index of $1/3$. If the last player splits up into two players, the new game is $[5; 2, 2, 1, 1]$. Each of the new players is pivotal for exactly one coalition, while each of the players of weight 2 is pivotal for three coalitions. Thus, the new players have a Banzhaf index of $1/8$ each. Similarly, each of the new players is pivotal for a permutation if and only if it appears in the third position, followed by the other new player, i.e., the new players have Shapley–Shubik index of $2/24 = 1/12$. Thus, weight splitting decreases the player's power by a factor of $4/3$ according to the Banzhaf index and by a factor of 2 according to the Shapley–Shubik index.

Finally, weight-splitting may have no effect on the player's power.

**Example 3. [Neutral splitting]** Consider the WVG $[4; 2, 2, 2]$. As in the previous examples, by symmetry, each player has a Banzhaf index of $1/3$, and a Shapley–Shubik index of $1/3$. If the last player splits up into two players, the new game is $[4; 2, 2, 1, 1]$. In this game, each of the new players is pivotal for 2 coalitions, while each of the players of weight 2 is pivotal for 4 coalitions. Thus, the split-up players have a Banzhaf index of $1/6$ each. Similarly, each of the new players is pivotal for a permutation if and only if it appears in the third position, followed by one of the players of weight 2. There are exactly 4 such permutations, so the Shapley–Shubik index of each of the new players is $1/6$. We have $2 \cdot 1/6 = 1/3$, i.e., according to both indices, the player's total power did not change.

In all examples presented so far, weight-splitting had the same effect on the Shapley–Shubik index and the Banzhaf index of the manipulator. We will now show that this is not always the case.

**Example 4.** Consider the WVG $[5; 2, 1, 1, 1, 1]$. In this game, the first player is pivotal for a permutation if he appears in the last or second-to-last position, but not in earlier positions.

Thus, his Shapley–Shubik index is 2/5. Further, this player is pivotal for any coalition that contains three or four players of weight 1, i.e., for 5 coalitions. On the other hand, any player of weight 1 is pivotal for any coalition that contains the player of weight 2 as well as any two other players of weight 1, i.e., for 3 coalitions. Thus the Banzhaf index of the first player is given by $5/(5 + 4 \cdot 3) = 5/17$.

Now, if the first player splits into two players of weight one, in the resulting game all players have the same weight. Therefore, for each of them the value of both indices is 1/6, and hence the total power of the manipulator is 1/3.

It remains to observe that $2/5 > 1/3$, but $5/17 < 1/3$, i.e., weight-splitting hurts the manipulator if the payoff is distributed according to the Shapley–Shubik index, but helps him if the Banzhaf index is used. Further, this example can be generalized to any weighted voting game of the form $[n; 2, 1, \ldots, 1]$, where there are $n - 1$ players of weight 1 and $n \geq 5$: in any such game, weight-splitting lowers the payoff of the first player according to the Shapley–Shubik index from $\frac{2}{n}$ to $\frac{2}{n+1}$, but increases his payoff according to the Banzhaf index from $\frac{n}{(n-1)^2+1}$ to $\frac{2}{n+1}$.

## 4. Splitting: Bounds of Manipulation

We have seen that a player can both increase and decrease his total payoff by splitting his weight. In this subsection, we provide upper and lower bounds on how much he can change his payoff by doing so. We restrict our attention to the case of splitting into two identities; the general case is briefly discussed in Section 9.

To simplify notation, in the rest of this section we assume that in the original game $G = [q; w_1, \ldots, w_n]$ the manipulator is player $n$, and he splits into two new identities $n'$ and $n''$, resulting in a new game $G'$. We first consider the case of the Shapley–Shubik index, followed by the analysis for the Banzhaf index.

### 4.1 Shapley–Shubik Index

We start by providing a tight upper bound on the benefits of manipulation.

**Theorem 5.** *For any game $G = [q; w_1, \ldots, w_n]$ and any split of $n$ into $n'$ and $n''$, we have $\varphi_{n'}(G') + \varphi_{n''}(G') \leq \frac{2n}{n+1}\varphi_n(G)$, i.e., the manipulator cannot gain more than a factor of $2n/(n+1) < 2$ by splitting his weight between two identities. Moreover, this bound is tight, i.e., there exists a game in which player $n$ increases his payoff by a factor of $2n/(n+1)$ by splitting into two identities.*

*Proof.* Fix a split of $n$ into $n'$ and $n''$. Let $\Pi_{n-1}$ be the set of all permutations of the first $n - 1$ players. Consider any $\pi \in \Pi_{n-1}$. Let $P(\pi)$ be the set of all permutations of the players in $G'$ that can be obtained by inserting $n'$ and $n''$ into $\pi$. Let $\Pi^*_{n+1}$ be the set of all permutations $\pi^*$ of players in $G'$ such that $n'$ or $n''$ is pivotal for $\pi^*$. Finally, let $P^*(\pi, k)$ be the subset of $P(\pi) \cap \Pi^*_{n+1}$ that consists of all permutations $\pi' \in P(\pi)$ in which at least one of the players $n'$ and $n''$ appears between the $k$-th and the $(k+1)$-st element of $\pi'$ and is pivotal for $\pi'$. Every permutation in $\Pi^*_{n+1}$ appears in one of the sets $P^*(\pi, k)$ for some

$\pi, k$, so we have

$$\varphi_{n'}(G') + \varphi_{n''}(G') = \frac{|\Pi_{n+1}^*|}{(n+1)!} \leq \frac{1}{(n+1)!} \sum_{\pi,k} |P^*(\pi,k)|.$$

On the other hand, it is not hard to see that $|P^*(\pi,k)| \leq 2n$ for any $\pi,k$: there are two ways to place $n'$ and $n''$ between the $k$-th and the $(k+1)$-st element of $\pi$, $n-1$ permutations in $P^*(\pi,k)$ in which $n'$ appears after the $k$-th element of $\pi$, but $n''$ is not adjacent to it, and $n-1$ permutations in $P^*(\pi,k)$ in which $n''$ appears after the $k$-th element of $\pi$, but $n'$ is not adjacent to it. Moreover, if $P^*(\pi,k)$ is not empty, then $n$ is pivotal for the permutation $f(\pi,k)$ obtained from $\pi$ by inserting $n$ after the $k$-th element of $\pi$. Further, if $(\pi_1, k_1) \neq (\pi_2, k_2)$ then $f(\pi_1, k_1) \neq f(\pi_2, k_2)$. Hence,

$$\varphi_n(G) \geq \frac{1}{n!} \sum_{\pi,k : P^*(\pi,k) \neq \emptyset} 1 \geq \frac{1}{n! \cdot 2n} \sum_{\pi,k} |P^*(\pi,k)| \geq \frac{n+1}{2n}(\varphi_{n'}(G') + \varphi_{n''}(G')).$$

We conclude that $\varphi_{n'}(G') + \varphi_{n''}(G') \leq \frac{2n}{n+1}\varphi_n(G) \leq 2\varphi_n(G)$, i.e., the manipulator cannot gain more than a factor of $2n/(n+1) < 2$ by splitting his weight between two identities.

To see that this bound is tight, consider the game $G = [2n; 2, 2, \ldots, 2]$ and suppose that one of the players (say, $n$) decides to split into two identities $n'$ and $n''$ resulting in the game $G' = [2n; 2, \ldots, 2, 1, 1]$. In both games the only winning coalition consists of all players, so we have $\varphi_n(G) = 1/n$, $\varphi_{n'}(G') = \varphi_{n''}(G') = 1/(n+1)$, i.e., $\varphi_{n'}(G') + \varphi_{n''}(G') = \frac{2n}{n+1}\varphi_n(G)$. □

We have seen that no player can increase his payoff by more than a factor of 2 by splitting his weight between two identities. In contrast, we will now show that a player can decrease his payoff by a factor of $\Theta(n)$ by doing so. This shows that a would-be manipulator has to be careful when deciding whether to split his weight, and motivates the algorithmic questions studied in the next two sections.

**Theorem 6.** *For any game $G = [q; w_1, \ldots, w_n]$ and any split of $n$ into $n'$ and $n''$, we have $\varphi_{n'}(G') + \varphi_{n''}(G') \geq \frac{n+1}{2}\varphi_n(G)$, i.e., the manipulator cannot lose more than a factor of $(n+1)/2$ by splitting his weight between two identities. Moreover, this bound is tight,*

*Proof.* To prove the first part of the theorem, fix a split of $n$ into $n'$ and $n''$ and consider any permutation $\pi$ of players in $G$ such that $n$ is pivotal for $\pi$. It is easy to see that at least one of $n'$ and $n''$ is pivotal for the permutation $f(\pi)$ obtained from $\pi$ by replacing $n$ with $n'$ and $n''$ (in this order). Similarly, at least one of $n'$ and $n''$ is pivotal for the permutation $g(\pi)$ obtained from $\pi$ by replacing $n$ with $n''$ and $n'$ (in this order). Moreover, all such permutations of players in $G'$ are distinct, i.e., for any $\pi, \pi'$ we have $g(\pi) \neq f(\pi')$, and $\pi \neq \pi'$ implies $f(\pi) \neq f(\pi')$, $g(\pi) \neq g(\pi')$. Hence, if $\Pi_n^*$ is the set of all permutations $\pi$ of the players in $G$ such that $n$ is pivotal for $\pi$, and $\Pi_{n+1}^*$ is the set of all permutations $\pi$ of the players in $G'$ such that $n'$ or $n''$ is pivotal for $\pi$, we have $|\Pi_{n+1}^*| \geq 2|\Pi_n^*|$ and

$$\varphi_{n'}(G') + \varphi_{n''}(G') = \frac{|\Pi_{n+1}^*|}{(n+1)!} \geq \frac{2|\Pi_n^*|}{(n+1)!} = \frac{2}{n+1}\varphi_n(G).$$

To see that this bound is tight, consider the game $G = [2n-1; 2, 2, \ldots, 2]$ and suppose that one of the players (say, $n$) decides to split into two identities $n'$ and $n''$ resulting in the game $G' = [2n - 1; 2, \ldots, 2, 1, 1]$. In the original game $G$, the only winning coalition consists of all players, so we have $\varphi_n(G) = 1/n$. Now, consider any permutation $\pi$ of the players in $G'$. We claim that $n'$ is pivotal for $\pi$ if and only if it appears in the $n$-th position of $\pi$, followed by $n''$. Indeed, if $\pi(n') = n$, $\pi(n'') = n + 1$, then all players in the first $n - 1$ positions have weight 2, so $w(S_\pi(n')) = 2n - 2$, $w(S_\pi(n') \cup \{n'\}) = 2n - 1$. Conversely, if $\pi(n') = n + 1$, we have $w(S_\pi(n')) = 2n - 1 = q$, and if $\pi(n') \leq n - 1$, we have $w(S_\pi(n') \cup \{n'\}) \leq 2(n - 1)$. Finally, if $\pi(n') = n$, but $\pi(n'') \neq n + 1$, we have $w(S_\pi(n') \cup \{n'\}) = 2n - 2 < q$. Consequently, $n'$ is pivotal for $(n-1)!$ permutations, and, by the same argument, $n''$ is also pivotal for (a disjoint set of) $(n - 1)!$ permutations. Hence, we have $\varphi_{n'}(G') + \varphi_{n''}(G') = \frac{2(n-1)!}{(n+1)!} = \frac{2}{n+1}\varphi_n(G)$. □

## 4.2 Banzhaf Index

For the Banzhaf index, we can obtain similar bounds on the maximum gains and losses from a weight-splitting manipulation.

**Theorem 7.** *For any game $G = [q; w_1, \ldots, w_n]$ and any split of $n$ into $n'$ and $n''$, we have $\beta_{n'}(G') + \beta_{n''}(G') \leq 2\beta_n(G)$. Moreover, this bound is asymptotically tight.*

*Proof.* Assume that player $n$ splits up into $n'$ and $n''$ and that $w_{n'} \leq w_{n''}$. Consider a losing coalition $C$ for which $n$ is critical in $G$. Then $w(C) < q \leq w(C) + w_n = w(C) + w_{n'} + w_{n''}$. We have the following possibilities:

- $q - w(C) \leq w_{n'}$. In this case $n'$ and $n''$ are critical for $C$ in $G'$.

- $w_{n'} < q - w(C) \leq w_{n''}$. In this case $n''$ is critical for $C \cup \{n'\}$ and $C$ in $G'$.

- $q - w(C) > w_{n''}$. In this case $n'$ is critical for $C \cup \{n''\}$ and $n''$ is critical for $C \cup \{n'\}$ in $G'$.

Therefore we have $\eta_{n'}(G') + \eta_{n''}(G') = 2\eta_n(G)$ in each case.

Now consider a player $i \in N \setminus \{n\}$. Suppose that $i$ is critical for a coalition $C$ in $G$. If $n \in C$, then $i$ is also critical for the coalition $C' = C \setminus \{n\} \cup \{n', n''\}$ in $G'$. On the other hand, if $n \notin C$, then $i$ also remains critical for $C$ in $G'$. Hence $\eta_i(G) \leq \eta_i(G')$. Moreover, $i$ may also be critical for some coalitions in $G'$ that contain just one of $n'$ and $n''$, so the above inequality will not in general be an equality. Thus, we have

$$\beta_{n'}(G') + \beta_{n''}(G') = \frac{2\eta_n(G)}{2\eta_n(G) + \sum_{i \in N \setminus \{n\}} \eta_i(G')}$$

$$\leq \frac{2\eta_n(G)}{2\eta_n(G) + \sum_{i \in N \setminus \{n\}} \eta_i(G)}$$

$$\leq \frac{2\eta_n(G)}{\eta_n(G) + \sum_{i \in N \setminus \{n\}} \eta_i(G)} = 2\beta_n(G).$$

To see that this bound is tight, consider a WVG $G = [n - 1; 1, \ldots, 1, 2]$ with $n$ players. We have $\eta_n(G) = n - 1 + \binom{n-1}{2}$ and $\eta_i(G) = 1 + \binom{n-2}{2}$ for $i \neq n$. Therefore,

$$\beta_n(G) = \frac{n - 1 + \binom{n-1}{2}}{n - 1 + \binom{n-1}{2} + (n-1)(1 + \binom{n-2}{2})} = \frac{n}{n^2 - 4n + 8} \sim 1/n.$$

If player $n$ splits up into two players $n'$ and $n''$ with weights 1 each, then in the resulting game $G'$ the Banzhaf index of each player is $\frac{1}{n+1}$. Thus for large $n$, $\beta_{n'}(G') + \beta_{n''}(G') = \frac{2}{n+1} \sim 2\beta_n(G)$. □

We can also bound the damage that can be incurred by weight-splitting.

**Theorem 8.** *For any game $G = [q; w_1, \ldots, w_n]$ and any split of $n$ into $n'$ and $n''$, we have $\beta_{n'}(G') + \beta_{n''}(G') \geq \frac{1}{n}\beta_n(G)$.*

*Proof.* Suppose that player $n$ splits into two players $n'$ and $n''$ with weights $w_{n'}$ and $w_{n''}$, respectively. We assume without loss of generality that $w_{n'} \leq w_{n''}$. Now, consider an arbitrary player $i \neq n$, and let

$$\mathcal{T}_i = \{S \subseteq N \setminus \{i\} \mid w(S) < q, w(S) + w_i \geq q\},$$
$$\mathcal{S}_i = \{S \subseteq N \setminus \{n, i\} \cup \{n', n''\} \mid w(S) < q, w(S) + w_i \geq q\}.$$

We have $\eta_i(G) = |\mathcal{T}_i|$, $\eta_i(G') = |\mathcal{S}_i|$. Further, set

$$\mathcal{S}_i^1 = \{S \in \mathcal{S}_i \mid i \text{ is pivotal for } S \setminus \{n', n''\}\},$$
$$\mathcal{S}_i^2 = \{S \in \mathcal{S}_i \mid i \text{ is not pivotal for } S \setminus \{n', n''\} \text{ and } n' \in S, n'' \notin S\},$$
$$\mathcal{S}_i^3 = \{S \in \mathcal{S}_i \mid i \text{ is not pivotal for } S \setminus \{n', n''\} \text{ and } n'' \in S, n' \notin S\},$$
$$\mathcal{S}_i^4 = \{S \in \mathcal{S}_i \mid i \text{ is not pivotal for } S \setminus \{n', n''\} \text{ and } n', n'' \in S\}.$$

We claim that $\mathcal{S}_i = \cup_{j=1}^4 \mathcal{S}_i^j$. Indeed, if $S \notin \mathcal{S}_i^1$, then $i$ is pivotal for $S$, but not for $S \setminus \{n', n''\}$, and hence it must be the case that $S \cap \{n', n''\} \neq \emptyset$; all such sets are included in $\mathcal{S}_i^1 \cup \mathcal{S}_i^2 \cup \mathcal{S}_i^3$.

For any $S \in \mathcal{S}_i$, let $f(S) = S \setminus \{n', n''\}$, $g(S) = S \setminus \{n', n''\} \cup \{n\}$. If $S \in \mathcal{S}_i^1$, then $f(S) \in \mathcal{T}_i$, and for each set $T \in \mathcal{T}_i$ there are at most 4 sets $S$ such that $f(S) = T$, i.e., $|f^{-1}(T)| \leq 4$. Further, $S \in \mathcal{S}_i^4$ implies $g(S) \in \mathcal{T}_i$, and $|g^{-1}(T)| \leq 1$ for any $T \in \mathcal{T}_i$. Finally, we have $g(S') \neq f(S'')$ for any $S', S'' \in \mathcal{S}_i$. Taken together, these observations imply that $|\mathcal{S}_i^1| + |\mathcal{S}_i^4| \leq 4|\mathcal{T}_i| = 4\eta_i(G)$.

Now, consider an $S \in \mathcal{S}_i^2$. We have $w(S) < q$, $w(S) + w_i < q$, $w(S) + w_i + w_{n'} \geq q$. Hence, $n'$ is critical for $S \cup \{i\}$. Similarly, if $S \in \mathcal{S}_i^3$, it follows that $n''$ is critical for $S \cup \{i\}$. Therefore, we have $|\mathcal{S}_i^2| + |\mathcal{S}_i^3| \leq \eta_{n'}(G') + \eta_{n''}(G')$. We obtain

$$\eta_i(G') = |\mathcal{S}_i| \leq \sum_{j=1}^4 |\mathcal{S}_i^j| \leq 4\eta_i(G) + \eta_{n'}(G') + \eta_{n''}(G') = 4\eta_i(G) + 2\eta_n(G),$$

where the last equality follows from the proof of Theorem 7. Thus, we obtain

$$\begin{aligned}
\beta_{n'}(G') + \beta_{n''}(G') &= \frac{2\eta_n(G)}{2\eta_n(G) + \sum_{i \in N \setminus \{n\}} \eta_i(G')} \\
&\geq \frac{2\eta_n(G)}{2\eta_n(G) + 4\sum_{i \in N \setminus \{n\}} \eta_i(G) + 2(n-1)\eta_n(G)} \\
&\geq \frac{2\eta_n(G)}{2n\sum_{i \in N} \eta_i(G)} = \frac{\beta_i(G)}{n}.
\end{aligned}$$

□

While it is not clear if the bound given in Theorem 8 is tight, our next example shows that splitting into two players can decrease a player's payoff according to the Banzhaf index by a factor of almost $\sqrt{\frac{n}{2\pi}}$.

**Example 9.** Consider a WVG $G = [3k; 1, \ldots, 1, 4k]$ with $n = 2k$ players. Let $N_1$ be the set of all players of weight 1, i.e., $N_1 = \{1, \ldots, n-1\}$. It is easy to see that player $n$ is critical for any coalition, while all other players are dummies, so we have $\beta_n(G) = 1$. Now suppose that player $n$ splits up into new identities $n'$ and $n''$ with weights $w_{n'} = w_{n''} = 2k$. For player $n'$ to be critical for a coalition $S$ in $G'$, it has to be the case that either $n'' \notin S$, $k \leq |S \cap N_1| \leq n-1$ or $n'' \in S$, $0 \leq |S \cap N_1| \leq k-1$. Thus, we have

$$\eta_{n'}(G') = \eta_{n''}(G') = \sum_{i=0}^{n} \binom{n-1}{i} = 2^{n-1}.$$

Moreover, for a player $i$ with weight 1 to be critical for a coalition in $G'$, the coalition must include exactly one of $n'$ or $n''$ as well as $k-1$ of the $n-2$ other players in $N_1 \setminus \{i\}$. Thus, we have $\eta_i(G') = 2\binom{2k-2}{k-1}$ for $i < n$. Using the standard formulas for the asymptotics of the central binomial coefficient, we can approximate $2\binom{2k-2}{k-1}$ by $2\sqrt{\frac{2}{\pi(2k-1)}}4^{k-1}$. We obtain

$$\beta_{n'}(G') + \beta_{n''}(G') \approx \frac{2 \cdot 2^{n-1}}{2^{n-1} + 2^{n-1} + (n-1)\sqrt{\frac{2}{\pi(n-1)}}2^{n-1}} = \frac{2}{2 + \sqrt{n-1}\sqrt{\frac{2}{\pi}}} \sim \sqrt{\frac{2\pi}{n}}.$$

## 5. Complexity of Finding a Beneficial Split

We now examine the problem of finding a beneficial weight split in weighted voting games from the computational perspective. Ideally, the manipulator would like to find a payoff-maximizing split, i.e., a way to split his weight among two or more identities that results in the maximal total payoff. A less ambitious goal is to decide whether there exists a manipulation that increases the manipulator's payoff. However, it turns out that even this problem is computationally hard. In the rest of the section, we show that checking whether there exists a payoff-increasing split is NP-hard both for the Shapley–Shubik index and the Banzhaf index; this holds even if the player is only allowed to use two identities. That is, in the spirit of the groundbreaking papers of Bartholdi and Orlin (1991) and Bartholdi, Tovey, and Trick (1989, 1992), we show that computational complexity acts as a barrier to manipulative behavior.

To formally define our computational problems, we require that all weights and the quota of both the original game and the new game are integers given in binary, i.e., we only allow integer splits. We remark that this assumption is not entirely without loss of generality: there are games where a player can benefit from a fractional split more than from any integer split. One such example is given by the game $[3; 1, 1, 1]$, where there are no non-trivial integer splits available to the players, but, similarly to Example 1, each player can increase his power by a factor of $3/2$ by splitting into two players with weight $1/2$. However, in real-life settings there is usually a natural bound on the granularity of the weights: if the weight is the number of supporters of a given party, it needs to be an integer, and if it is the monetary contribution of a player, it usually has to be an integer number of dollars (or, at least, cents), i.e., our assumption reflects real-life constraints.

We are now ready to define our problems.

**Name**: Beneficial-SS-Split
**Instance**: $(G, \ell)$ where $G = [q; w_1, \ldots, w_n]$ is a weighted voting game and $\ell \in \{1, \ldots, n\}$.
**Question**: Is there a way for player $\ell$ to split his weight $w_\ell$ between sub-players $\ell_1, \ldots, \ell_m$ so that in the new game $G'$ it holds that $\sum_{j=1}^m \varphi_{\ell_j}(G') > \varphi_\ell(G)$?

The definition of Beneficial-BI-Split is similar.

**Name**: Beneficial-BI-Split
**Instance**: $(G, \ell)$ where $G = [q; w_1, \ldots, w_n]$ is a weighted voting game and $\ell \in \{1, \ldots, n\}$.
**Question**: Is there a way for player $\ell$ to split his weight $w_\ell$ between sub-players $\ell_1, \ldots, \ell_m$ so that in the new game $G'$ it holds that $\sum_{j=1}^m \beta_{\ell_j}(G') > \beta_\ell(G)$?

Note that we are looking for a *strictly* beneficial manipulation, i.e., one that *increases* the manipulator's total payoff, rather than one that is simply *not harmful*.

We will prove that Beneficial-SS-Split and Beneficial-BI-Split are NP-hard. Our hardness results are based on reductions from the following classic NP-hard problem:

**Name**: Partition
**Instance**: A set of $k$ integer weights $A = \{a_1, \ldots, a_k\}$.
**Question**: Is it possible to partition $A$ into two subsets $P_1 \subseteq A$, $P_2 \subseteq A$ so that $P_1 \cap P_2 = \emptyset$, $P_1 \cup P_2 = A$, and $\sum_{a_i \in P_1} a_i = \sum_{a_i \in P_2} a_i$?

We will first prove a simple lemma that will be used in all NP-hardness proofs in this paper.

**Lemma 10.** *Let $A = \{a_1, \ldots, a_k\}$ be a "no"-instance of* Partition. *Then for any weighted voting game $G = [q; w_1, \ldots, w_n]$ such that $n > k$, $w_i = 8a_i$ for $i = 1, \ldots, k$, $q = 4\sum_{a_i \in A} a_i + r$, where $0 < r < 4$, and $\sum_{i=k+1}^n w_i < 4$, it holds that all players $k+1, \ldots, n$ are dummies, and hence their Shapley–Shubik and Banzhaf indices are equal to 0.*

*Proof.* Consider a player $i$ with $k < i \leq n$ and a set $S \subseteq N \setminus \{i\}$. We will show that $i$ is not pivotal for $S$.

Set $N_0 = \{1, \ldots, k\}$ and let $S_0 = S \cap N_0$. The set $N_0$ can be partitioned into two equal-weight subsets if and only if $A$ can, so either $w(S_0) < w(N_0)/2$, or $w(S_0) > w(N_0)/2$. Moreover, the weights of all players in $N_0$ are multiples of 8, so $w(N_0)/2$ is a multiple of 4. Similarly, the weight of $S_0$ is a multiple of 8. Hence, if $w(S_0) < w(N_0)/2$, it follows that $w(S_0) \leq w(N_0)/2 - 4$ and $w(S \cup \{i\}) < w(N_0)/2 - 4 + 4 < q$. Therefore, we have $v(S) = 0$, $v(S \cup \{i\}) = 0$, i.e., $i$ is not pivotal for $S$. On the other hand, if $w(S_0) > w(N_0)/2$, then $w(S_0) \geq w(N_0)/2 + 4 > q$, so $S_0$ is a winning coalition. Therefore, $i$ is not pivotal for $S$ in this case as well. $\square$

**Theorem 11.** Beneficial-BI-Split *is NP-hard, and remains NP-hard even if the player can only split into two players with equal weights.*

*Proof.* Given an instance of Partition $A = \{a_1, \ldots, a_k\}$, we construct a weighted voting game $G = [q; w_1, \ldots, w_n]$ with $n = k + 1$ players as follows. We let $X = \sum_{a_i \in A} a_i$, and set $w_i = 8a_i$ for $i = 1, \ldots, n-1$, $w_n = 2$, and $q = 4X + 2$. Also, we set $\ell = n$. Since $w_n = 2$, the only integer split available to player $n$ is into two identities $n'$ and $n''$ with weight 1 each. Let $G' = [q; w_1, \ldots, w_{n-1}, 1, 1]$ be the resulting game.

If $A$ is a "no"-instance of Partition, then Lemma 10 implies that player $n$ is a dummy, and, moreover, if he splits into sub-players, these sub-players are also dummies. Therefore $(G, \ell)$ is a "no"-instance of Beneficial-Bi-Split.

Now let us assume that $A$ is a "yes"-instance of Partition. Let $x$ denote the number of coalitions in $N \setminus \{n\}$ of weight $4X$. Then $\eta_n(G) = x$. For $i = 1, \ldots, n-1$, let

$$\mathcal{S}_i = \{S \subseteq N \setminus \{n, i\} \mid w(S) < 4X, w(S) + w_i \geq q\},$$

and set $y_i = |\mathcal{S}_i|$. Also, set $y = \sum_{i=1}^{n-1} y_i$.

Consider a player $i < n$. Observe that exactly half of the $x$ subsets of $\{1, \ldots, n-1\}$ of weight $4X$ contain $i$. For any such subset $T$, player $i$ is pivotal for $(T \setminus \{i\}) \cup \{n\}$. Further, for any coalition $S \in \mathcal{S}_i$, player $i$ is pivotal for both $S$ and $S \cup \{n\}$. Therefore for $i < n$ we have $\eta_i(G) = \frac{x}{2} + 2y_i$. We obtain

$$\beta_n(G) = \frac{x}{x + (n-1)\frac{x}{2} + 2y}.$$

On the other hand, in the new game $G'$ we have $\eta_{n'}(G') = \eta_{n''}(G') = x$. Moreover, for $i < n$ we have $\eta_i(G') = \frac{x}{2} + 4y_i$, since each coalition in $\mathcal{S}_i$ corresponds to 4 coalitions for which $i$ is pivotal, namely, $S$, $S \cup \{n'\}$, $S \cup \{n''\}$, and $S \cup \{n', n''\}$. Thus,

$$\beta_{n'}(G') + \beta_{n''}(G') = \frac{2x}{2x + (n-1)\frac{x}{2} + 4y} > \beta_n(G),$$

where the last inequality holds since $x > 0$. Thus, a "yes"-instance of Partition corresponds to a "yes"-instance of Beneficial-Bi-Split. $\qquad \square$

We now consider the problem of finding a beneficial split for the Shapley–Shubik index.

**Theorem 12.** Beneficial-SS-Split *is NP-hard, and remains NP-hard even if the player can only split into two players with equal weights.*

*Proof.* Given an instance $A = \{a_1, \ldots, a_k\}$ of Partition, we set $X = \sum_{a_i \in A} a_i$, and create a weighted voting game $G = [4X + 3; 8a_1, \ldots, 8a_k, 1, 2]$ with $n = k + 2$ players. Also, we set $N_0 = \{1, \ldots, n-2\}$.

If $A$ is a "no"-instance of Partition, then Lemma 10 implies that player $n$ is a dummy, and if he splits into several players, all of them will be dummies. Thus, we have constructed a "no"-instance of Beneficial SS-Split.

Now, suppose that $A$ is a "yes"-instance of Partition. Let $\langle P_1, P_2 \rangle$ be a partition of $A$, so $w(P_1) = w(P_2)$. It corresponds to a partition $\langle S, N_0 \setminus S \rangle$ of $N_0$, where $i \in S$ if and only if $a_i \in P_1$; observe that $w(S) = w(N_0 \setminus S)$. Set $s = |S|$, so $|N_0 \setminus S| = n - s - 2$.

It is easy to see that $n$ is critical for $S \cup \{n-1\}$ as well as for $(N_0 \setminus S) \cup \{n-1\}$. There are $(s+1)!(n-2-s)!$ permutations of $1, \ldots, n$ that put $n$ directly after some permutation of $S \cup \{n-1\}$. Similarly, there are $s!(n-1-s)!$ permutations putting $n$ directly after some permutation of $(N_0 \setminus S) \cup \{n-1\}$. Thus, for each partition $P^i = \langle P_1^i, P_2^i \rangle$, where $|P_1^i| = s$, we have at least $(s+1)!(n-2-s)! + s!(n-1-s)!$ distinct permutations where $n$ is critical. On the other hand, as argued above, if $S$ is a subset of $N_0$ such that $w(S) \neq w(N_0)/2$, then $n$ is not critical for $S$ or $S \cup \{n-1\}$, since either $w(S) \leq w(N_0)/2 - 4 < q - 3$ or $w(S) \geq w(N_0)/2 + 4 > q$.

Let $\mathcal{P}$ be the set of all partitions of $A$, where each partition is counted only once, i.e., $\mathcal{P}$ contains exactly one of the $\langle P_1, P_2 \rangle$ and $\langle P_2, P_1 \rangle$. For each $P^i = \langle P_1^i, P_2^i \rangle \in \mathcal{P}$, we denote $|P_1^i| = s_i$. There is a total of $n!$ permutations of players in $G$. Thus, the Shapley–Shubik index of $n$ in $G$ is

$$\varphi_n(G) = \sum_{P^i \in \mathcal{P}} \frac{(s_i+1)!(n-2-s_i)! + s_i!(n-1-s_i)!}{n!} =$$

$$\sum_{P^i \in \mathcal{P}} \frac{s_i!(n-2-s_i)!(s_i+1+n-1-s_i)}{n!} =$$

$$\sum_{P^i \in \mathcal{P}} \frac{n s_i!(n-2-s_i)!}{n!} = \sum_{P^i \in \mathcal{P}} \frac{s_i!(n-2-s_i)!}{(n-1)!}.$$

We now consider what happens when $n$ splits into two players, $n'$ and $n''$ with $w_{n'} = w_{n''} = 1$, resulting in a game $G' = [4X + 3; 8a_1, \ldots, 8a_k, 1, 1, 1]$.

Again, let $\langle P_1, P_2 \rangle$, $|P_1| = s_i$, $|P_2| = n - s_i$, be a partition of $A$ such that $w(P_1) = w(P_2)$, and let $\langle S, N_0 \setminus S \rangle$ be the corresponding partition of $N_0$. There are $(s_i+2)!(n-2-s_i)!$ permutations that place $n''$ directly after some permutation of $S \cup \{n-1, n'\}$, and $n''$ is critical for each of them. Similarly, $n''$ is critical for the $s_i!(n-s_i)!$ permutations that place $n''$ directly after some permutation of $(N_0 \setminus S) \cup \{n-1, n'\}$.

Thus, each partition $P^i = \langle P_1^i, P_2^i \rangle$ with $|P_1^i| = s_i$, corresponds to $(s_i+2)!(n-2-s_i)! + s_i!(n-s_i)!$ distinct permutations where $n''$ is critical. By symmetry, there are $(s_i+2)!(n-2-s_i)! + s_i!(n-s_i)!$ distinct permutations where $n'$ is critical. There are $n+1$ players in $G'$, so there is a total of $(n+1)!$ permutations of the players. Thus each partition $P^i = \langle P_1^i, P_2^i \rangle$, $|P_1^i| = s_i$, contributes $\frac{s_i!(n-2-s_i)!}{(n-1)!}$ to the Shapley–Shubik index of $n$ in $G$, and $2 \frac{(s_i+2)!(n-2-s_i)! + s_i!(n-s_i)!}{(n+1)!}$ to the sum of the Shapley–Shubik indices of $n'$ and $n''$ in $G'$. We will now show that for any partition $P^i$

$$2 \frac{(s_i+2)!(n-2-s_i)! + s_i!(n-s_i)!}{(n+1)!} > \frac{s_i!(n-2-s_i)!}{(n-1)!}. \tag{2}$$

Summing these inequalities over all partitions $P^i$ implies $\varphi_{n'}(G') + \varphi_{n''}(G') > \varphi_n(G)$, as desired. To prove inequality (2), note that it can be simplified to

$$2 \frac{(s+1)(s+2) + (n-1-s)(n-s)}{n(n+1)} > 1,$$

where we use $s$ instead of $s_i$ to simplify notation, or, equivalently,

$$2(s+1)(s+2) + 2(n-1-s)(n-s) - n(n+1) > 0.$$

Now, observe that

$$2(s+1)(s+2) + 2(n-1-s)(n-s) - n(n+1) = (n-2-2s)^2 + n > 0$$

for any $n > 0$. This proves inequality (2) for any $n > 0$. It follows that if $A$ is a "yes"-instance of PARTITION, player $n$ always gains by splitting into two players of weight 1, i.e., $(G, n)$ is a "yes"-instance of BENEFICIAL-SS-SPLIT. $\square$

**Remark 13.** It can be verified that both of our proofs go through even if we allow non-integer splits, i.e., our hardness results are independent of the integrality assumption. Further, note that we have not shown that BENEFICIAL-BI-SPLIT and BENEFICIAL-SS-SPLIT are in NP, i.e., we have not proved that these problems are NP-complete. There are two reasons for this. First, if we allow splits into an arbitrary number of identities, some of the candidate solutions may have exponentially many new players (e.g., a player with weight $w_i$ can split into $w_i$ players of weight 1). Second, even if we circumvent this issue by only considering splits into a polynomial number of identities, it is not clear how to verify in polynomial time whether a particular split is beneficial. In fact, since computing both power indices in weighted voting games is #P-hard, it is quite possible that our problems are not in NP.

## 6. Computing Beneficial Splits

In Section 5, we have shown that it is hard even to test if any beneficial split exists, let alone to find the optimal split. This can be seen as a positive result, since complexity of finding beneficial splits serves as a barrier for this kind of manipulative behavior. However, it turns out that in many cases manipulators can overcome the problem. More precisely, in what follows we show that in certain restricted domains manipulators can find beneficial splits into two identities in polynomial time. We then consider manipulation algorithms that work by approximating the Shapley–Shubik index (rather than calculating it precisely).

### 6.1 Examples

In this subsection, we describe two scenarios in which one of the players can always increase his payoff by splitting. While both of our examples rely on rather severe constraints on the players' weights and the threshold, there are practical weighted voting scenarios that satisfy these constraints.

**Example 14.** It is not hard to see that Example 1 can be generalized to any weighted voting game with $w(N) = q$; such games are sometimes called *unanimity games*. Even more generally, a player can always increase his payoff by weight-splitting if the threshold is set so high that any winning coalition must include all players; this holds both for the Shapley–Shubik index and the Banzhaf index. Indeed, consider the class of weighted voting games $G = [q; \mathbf{w}]$ that is characterized by the following condition: $w(N) - s < q \leq w(N)$, where $s = \min\{\min_i w_i, \lfloor w_{\max}/2 \rfloor\}$. The condition $s \leq \min_i w_i$ ensures that all players are present in all winning coalitions, so the index value of each player is $1/n$. Now, suppose that the player $i$ with the largest weight $w_i = w_{\max}$ splits his weight (almost) equally between two identities, i.e., sets $w_i' = \lfloor w_i/2 \rfloor$, $w_i'' = \lceil w_i/2 \rceil$. As we also have $s \leq \lfloor w_{\max}/2 \rfloor$, any winning coalition still has to include all players. Therefore, in the new game the payoff of each player is $1/(n + 1)$, and hence the split increases the total payoff of the manipulator by a factor of $2n/(n + 1)$.

**Example 15.** Our second example is specific to the Shapley–Shubik index. In this example, a "small" player can benefit from manipulation in the presence of "large" players, as long as the threshold is sufficiently high. Formally, consider the class of weighted voting games of the form $G = [q; \mathbf{w}]$, where all $w_i$, $i = 1, \ldots, n-1$, are multiples of some integer $A$, the threshold $q$ is of the form $AT + b$, $b < A$, and $b < w_n < \min\{2b - 1, A\}$. Suppose that all winning coalitions have size at least $\lceil n/2 \rceil + 1$. This condition holds if we renumber the players so that $w_1 \geq w_2 \geq \cdots \geq w_{n-1}$ and require $q > \sum_{i=1,\ldots,\lceil n/2 \rceil} w_i$.

Now, suppose that player $n$ is pivotal for at least one coalition in this game (if all weights are small multiples of $A$, this condition can be checked easily). Consider any permutation $\pi$ such that $n$ is pivotal for $\pi$. We have $w(S_\pi(n)) = AT$. Indeed, if $w(S_\pi(n)) > AT$, then $w(S_\pi(n)) \geq AT + A > q$, and the coalition $S_\pi(n)$ does not need player $n$ to win. On the other hand, if $w(S_\pi(n)) < AT$, then $w(S_\pi(n)) \leq AT - A$, so $w(S_\pi(n)) + w_n < AT + b = q$. Let $P$ be the set of all such permutations; we have $\varphi_n(G) = \frac{|P|}{n!}$.

Now suppose that $n$ decides to split its weight between two new identities $n'$ and $n''$ by setting $w'_n = b - 1$, $w''_n = w_n - b + 1$; note that $w'_n, w''_n < b$. Consider any permutation $\pi$. Suppose that $n$ occurs in the $k$-th position in this permutation. By our assumption, $k \geq \lceil n/2 \rceil + 1$. We will construct $2k$ permutations $\pi'_j$, $\pi''_j$, $j = 1, \ldots, k$, as follows. In each of these permutations, players $1, \ldots, n-1$ appear in the same order as in $\pi$. Moreover, in $\pi'_j$, player $n'$ occurs in the $j$-th position, and player $n''$ occurs in the $(k+1)$-st position. Similarly, in $\pi''_j$, player $n''$ occurs in the $j$-th position, and player $n'$ occurs in the $(k+1)$-st position.

Observe that $n''$ is pivotal for any $\pi'_j$, $j = 1, \ldots, k$. Indeed, the total weight of all players that precede $n''$ in $\pi'_j$ is $AT + w'_n < q$, while $w(S_{\pi'_j}(n'') \cup \{n''\}) > q$. Similarly, $w(S_{\pi''_j}(n')) < q$, so $n'$ is pivotal for any $\pi''_j$, $j = 1, \ldots, k$. Hence, the total number of permutations of $1, \ldots, n-1, n', n''$ for which either $n'$ or $n''$ is pivotal is at least $2k|P| \geq (n+2)|P|$, and the total Shapley–Shubik index of these players is at least $\frac{(n+2)|P|}{(n+1)!} > \frac{|P|}{n!} = \varphi_n(G)$. Hence, any such split is strictly beneficial for player $n$.

In addition to the scenarios discussed above, Fatima et al. (2007) describe several classes of voting games where the Shapley–Shubik indices of all players can be computed in polynomial time; Aziz and Paterson (2008) prove similar results for the Banzhaf index. Clearly, if the manipulator's weight is polynomially bounded, and the original game as well as all games that result from the manipulator splitting into two identities are "easy", i.e., belong to one the classes considered by Fatima et al. or Aziz and Paterson, the problem of finding the beneficial two-way split can be solved in polynomial time. However, our examples illustrate that a player may be able to decide whether it is beneficial to split even if he cannot compute his payoff prior to the manipulation.

## 6.2 Pseudopolynomial and Approximation Algorithms

The hardness reductions in Section 5 are from PARTITION. While this problem in known to be NP-hard, its hardness relies crucially on the fact that the weights of the elements are represented in binary. Indeed, if the weights are given in unary, there is a dynamic programming-based algorithm for this problem that runs in time polynomial in the size of the input (such algorithms are usually referred to as *pseudopolynomial*). In particular, if all weights are polynomial in $n$, the running time of this algorithm is polynomial in $n$. In many natural voting domains the weights of all players are not too large, so this scenario is quite realistic. It is therefore natural to ask if there exists a pseudopolynomial algorithm for the problem of finding a beneficial split.

It turns out that the answer to this question is indeed positive as long as there is a constant upper bound $K$ on the number of identities that the manipulator can use and all weights are required to be integers. To see this, recall that there is a pseudopolynomial algorithm for computing the Shapley–Shubik index of any player in a weighted voting game (Matsui & Matsui, 2000). This algorithm is based on dynamic programming: for any weight $W$ and any $1 \leq k \leq n$, it calculates the number of coalitions of size $k$ that have weight $W$. Thus, it can be easily adapted to work for the Banzhaf index as well.

One can use the algorithm of Matsui and Matsui (2000) to find a beneficial split for a player $i$ with weight $w_i$ in a game $G$ as follows. Consider all possible splits $w_i = w_i^{(1)} + \cdots + w_i^{(K)}$, where $w_i^{(j)} \in \mathbb{N}$ for $j = 1, \ldots, K$. The number of such splits is at most $(w_i)^K$, which is polynomial in $n$ for constant $K$. Evaluate the Shapley–Shubik indices (respectively, Banzhaf indices) of all new players in any such split and return "yes" if and only if at least one of these splits results in an increased total payoff. Let $A(G)$ be the running time of the algorithm of Matsui and Matsui on instance $G$. The running time of our algorithm is $O((w_i)^K K \cdot A(G))$, which is clearly pseudopolynomial.

We will now consider a more general setting, where only the weight of the manipulator is polynomially bounded, while the weights of other players can be large. To simplify the presentation, we limit ourselves to the case of two-way splits and the Shapley–Shubik index; however, our approach also applies to splits into any constant number of identities and to the Banzhaf index. We can use the same high-level approach as in the previous case, i.e., considering all possible splits (because of the weight restriction, there are only polynomially many of them), and computing the indices of both new players for each split. However, if we were to implement the latter step exactly, it would take exponential time. Therefore, in this version of our algorithm, we replace the algorithm of Matsui and Matsui with an approximation algorithm for computing the Shapley–Shubik index. Several such algorithms are known; see, e.g., the work of Mann and Shapley (1960), Fatima et al. (2007), Bachrach et al. (2010). We will use these algorithms in a black-box fashion. Namely, we assume that we are given a procedure $\texttt{Shapley}(G, i, \delta, \epsilon)$ that for any given values of $\epsilon > 0$ and $\delta > 0$ outputs a number $v$ that with probability $1 - \delta$ satisfies $|v - \varphi_i(G)| \leq \epsilon$ and runs in time $\text{poly}(n \log w_{\max}, 1/\epsilon, 1/\delta)$. We will now show how to use this procedure to design an algorithm for finding a beneficial split and relate the performance of our algorithm to that of $\texttt{Shapley}(G, i, \delta, \epsilon)$.

Our algorithm is given in Figure 1. It takes parameters $\delta$ and $\epsilon$ as inputs, and uses the procedure $\texttt{Shapley}(G, i, \delta, \epsilon)$ as a subroutine. The algorithm outputs "yes" if it finds a split whose total estimated payoff exceeds the payoff of the manipulator in the original game by at least $3\epsilon$. It can easily be modified to output the (approximately) optimal split.

**Proposition 16.** *With probability $1 - 3\delta$, the output of our algorithm satisfies the following: (i) If the algorithm outputs "yes", then $(G, i)$ admits a beneficial integer split; (ii) Conversely, if there is an integer split that increases the payoff to the manipulator by more than $6\epsilon$, our algorithm outputs "yes". Moreover, the running time of our algorithm is polynomial in $n w_i$, $1/\epsilon$, and $1/\delta$.*

```
FindSplit(G = [q; w], i, δ, ε);
v* =Shapley(G, i, δ, ε);
for j = 0, . . . , w_i
    w'_i = j, w''_i = w_i − j;
    G' = [q; w_1, . . . , w_{i−1}, w'_i, w''_i, w_{i+1}, . . . , w_n];
    v' = Shapley(G', i', δ, ε),  v'' =Shapley(G', i'', δ, ε);
    v = v' + v'';
    if v > v* + 3ε then return yes;
return no;
```

Figure 1: Algorithm $\mathtt{FindSplit}(G = [q; \mathbf{w}], i, \delta, \epsilon)$

*Proof.* Suppose that the algorithm outputs "yes". Consider the quantities $v^*$, $v'$ and $v''$ computed by our algorithm. We have $Prob[v^* < \varphi_i(G) − \epsilon] < \delta$, $Prob[v' > \varphi_{i'}(G') + \epsilon] < \delta$, $Prob[v'' > \varphi_{i''}(G') + \epsilon] < \delta$. Hence, with probability at least $1 − 3\delta$, if $v' + v'' > v^* + 3\epsilon$, then $\varphi_{i'}(G) + \varphi_{i''}(G') + 2\epsilon > \varphi_i(G) − \epsilon + 3\epsilon$, or, equivalently, $\varphi_{i'}(G') + \varphi_{i''}(G') > \varphi_i(G)$.

Conversely, suppose that there is a beneficial split of the form $(w'_i, w''_i)$ that improves player $i$'s payoff by at least $6\epsilon$. As before, with probability at least $1 − 3\delta$ we have that $v^* \le \varphi_i(G) + \epsilon$ and at the step $j = w'_i$ it holds that $v' \ge \varphi_{i'}(G') − \epsilon$, $v'' \ge \varphi_{i''}(G') − \epsilon$. Then $v = v' + v'' \ge \varphi_{i'}(G') + \varphi_{i''}(G') − 2\epsilon > \varphi_i(G) + 6\epsilon − 2\epsilon \ge v^* + 3\epsilon$, so the algorithm will output "yes". □

While our algorithm does not *guarantee* finding a successful manipulation, it is possible to control the approximation quality (at the cost of increasing the running time), so that a successful manipulation is found with high probability.

Thus we can see that manipulators have several ways to overcome the computational difficulty of finding the optimal manipulation. Hence, other measures are required to avoid such manipulations.

## 7. Merging and Annexation

Instead of a player splitting into smaller players, some players may merge into a single entity. However, this situation is very different from the game-theoretic perspective, as it involves coordinated actions by several would-be manipulators who then have to decide how to split the (increased) total payoff. For the case of players merging to gain advantage, we examine two cases. One is *annexation* where one player takes the voting weight of other players. The annexation is advantageous if the payoff of the new merged coalition in the new game is greater than the payoff of the annexer in the original game. The other case is *voluntary merging* where players merge to become a bloc so that their new payoff exceeds the sum of their individual payoffs.

For any weighted voting game $G$, we denote the game that results from merging of the players in a coalition $S$ by $G_{\&S}$; the set of players in the new game is $(N \setminus S) \cup \{\&S\}$, and its characteristic function is denoted by $v_{\&S}$. We will now define the computational problems of checking whether there exist a beneficial voluntary merge or annexation with respect to the Shapley–Shubik index.

**Name**: Beneficial-SS-Merge
**Instance**: $(G, S)$ where $G = [q; w_1, \ldots, w_n]$ is a weighted voting game and $S \subset \{1, \ldots, n\}$.
**Question**: If coalition $S$ merges to form a new game $G_{\&S}$, is $\varphi_{\&S}(G_{\&S}) > \sum_{i \in S} \varphi_i(G)$?

**Name**: Beneficial-SS-Annexation
**Instance**: $(G, S, i)$ where $G = [q; w_1, \ldots, w_n]$ is a weighted voting game, $1 \leq i \leq n$, and $S \subset \{1, \ldots, n\} \setminus \{i\}$.
**Question**: If $i$ annexes $S$ to form a new game $G_{\&(S \cup \{i\})}$, is $\varphi_{\&(S \cup \{i\})}(G_{\&(S \cup \{i\})}) > \varphi_i(G)$?

We can easily adapt these definitions for the Banzhaf index; we will refer to the resulting problems as Beneficial-BI-Merge and Beneficial-BI-Annexation. We will first consider the issues related to annexation, followed by the analysis for merging.

## 7.1 Merging

As in the case of splitting, we expect it to be hard to find a beneficial merge. The following theorem confirms this intuition.

**Theorem 17.** Beneficial-SS-Merge *and* Beneficial-BI-Merge *are NP-hard.*

*Proof.* Given an instance of Partition $A = \{a_1, \ldots, a_k\}$, we construct a weighted voting game $G = [q; w_1, \ldots, w_n]$ with $n = k + 3$ players as follows. We set $X = 4 \sum_{i=1}^{k} a_i$ and let $w_i = 8a_i$ for $i = 1, \ldots, n-3$, $w_{n-2} = w_{n-1} = w_n = 1$, and $q = X + 2$. We will now argue that $A$ is a "yes"-instance of Partition if and only if $(G, \{n-1, n\})$ is a "yes"-instance of both Beneficial-SS-Merge and Beneficial-BI-Merge.

If $A$ is a "no"-instance of Partition, then by Lemma 10 players $n$ and $n-1$ are dummies, and even if they merge together, the new player $\&\{n-1, n\}$ remains a dummy in the new game $G_{\&\{n-1,n\}}$. Thus, in this case $(G, \{n-1, n\})$ is a "no"-instance for both of our problems.

Now let us assume that $A$ is a "yes"-instance of Partition. We will first consider the case of the Shapley–Shubik index, followed by the analysis for the Banzhaf index.

Set $N_0 = \{1, \ldots, n-3\}$. Let $\langle P_1, P_2 \rangle$ be a partition of $A$, and let $\langle S, N_0 \setminus S \rangle$ be the corresponding partition of $N_0$. Set $s = |S|$, so $|N_0 \setminus S| = n - s - 3$. Player $n$ is critical for $S \cup \{n-2\}$ and $S \cup \{n-1\}$, as well as for $(N_0 \setminus S) \cup \{n-2\}$ and $(N_0 \setminus S) \cup \{n-1\}$. Thus, for each partition $P = \langle P_1, P_2 \rangle$, where $|P_1| = s$, we have exactly $4(s+1)!(n-2-s)!$ distinct permutations where $n$ is critical. Further, it is easy to see that $n$ is not critical for any other permutation. By symmetry, the same is true for $n-1$. Thus, each partition $\langle P_1, P_2 \rangle$ of $A$ with $|P_1| = s$ contributes $8 \frac{(s+1)!(n-2-s)!}{n!}$ to the sum of the Shapley–Shubik indices of $n-1$ and $n$.

Now, consider the game $G_{\&\{n-1,n\}}$. Consider a partition $\langle S, N_0 \setminus S \rangle$ of $N_0$ with $|S_0| = s$ that corresponds to a partition $\langle P_1, P_2 \rangle$ of $A$. Player $\&\{n-1, n\}$ is critical for $S$ and $S \cup \{n-2\}$, as well as for $N_0 \setminus S$ and $(N_0 \setminus S) \cup \{n-2\}$. Thus, each partition $\langle P_1, P_2 \rangle$ of $A$ with $|P_1| = s$ contributes $2 \frac{s!(n-2-s)!}{(n-1)!} + 2 \frac{(s+1)!(n-3-s)!}{(n-1)!}$ to the Shapley–Shubik index of

&$\{n-1, n\}$. It remains to show that

$$8\frac{(s+1)!(n-2-s)!}{n!} < 2\frac{s!(n-2-s)!}{(n-1)!} + 2\frac{(s+1)!(n-3-s)!}{(n-1)!}.$$

This inequality can be simplified to $4(s+1)(n-2-s) < n(n-2-s+s+1) = n(n-1)$, which is equivalent to $0 < n(n-1) - 4(s+1)(n-2-s) = (n-2s-3)^2 + n - 1$. This inequality clearly holds for $n \geq 1$. Hence, $\varphi_{n-1}(G) + \varphi_n(G) < \varphi_{\&\{n-1,n\}}(G_{\&\{n-1,n\}})$, i.e., $(G, \{n-1, n\})$ is a "yes"-instance of BENEFICIAL-SS-MERGE if and only if we started with a "yes"-instance of PARTITION.

We will now show that the same is true for the Banzhaf index. Let $x$ denote the number of coalitions in $N_0$ of weight $4X$. Then $\eta_{n-2}(G) = \eta_{n-1}(G) = \eta_n(G) = 2x$. For $i = 1, \ldots, k$, let

$$\mathcal{S}_i = \{S \subseteq N_0 \setminus \{i\} \mid w(S) < 4X, w(S) + w_i \geq q\},$$

and set $y_i = |\mathcal{S}_i|$. Also, set $y = \sum_{i=1}^k y_i$.

Consider a player $i \in N_0$. Observe that exactly half of the $x$ subsets of $N_0$ of weight $4X$ contain $i$. For any such set $T$, player $i$ is pivotal for $(T \setminus \{i\}) \cup \{n-2, n-1\}$, $(T \setminus \{i\}) \cup \{n-2, n\}$, $(T \setminus \{i\}) \cup \{n-1, n\}$, and $(T \setminus \{i\}) \cup \{n-2, n-1, n\}$. Further, for any coalition $S \in \mathcal{S}_i$, player $i$ is pivotal for any coalition of the form $S \cup S'$, where $S' \subseteq \{n-2, n-1, n\}$. Therefore for any $i \in N_0$ we have $\eta_i(G) = \frac{4x}{2} + 8y_i$. Thus, we have $\beta_{n-1}(G) = \beta_n(G) = \frac{2x}{6x+2kx+8y}$.

In the new game $G_{\&\{n-1,n\}}$, we have $\eta_{\&\{n-1,n\}}(G_{\&\{n-1,n\}}) = 2x$, but $\eta_{n-2}(G_{\&\{n-1,n\}}) = 0$. Now, consider a player $i \in N_0$ and a coalition $T \subset N_0$ of weight $4X$ that contains $i$. Player $i$ is pivotal for $(T \setminus \{i\}) \cup \{\&\{n-1, n\}\}$ and for $(T \setminus \{i\}) \cup \{n-2, \&\{n-1, n\}\}$. as well as for any coalition of the form $S \cup S'$, where $S \in \mathcal{S}_i$ and $S' \subseteq \{n-2, \&\{n-1, n\}\}$. Hence, $\eta_i(G_{\&\{n-1,n\}}) = \frac{2x}{2} + 4y_i$.

We obtain

$$\beta_{\&\{n-1,n\}}(G_{\&\{n-1,n\}}) = \frac{2x}{2x+kx+4y} > \frac{4x}{6x+2kx+8y} = \beta_{n-1}(G) + \beta_n(G).$$

Thus, $(G, \{n-1, n\})$ is a "yes"-instance of BENEFICIAL-BI-MERGE if and only if we started with a "yes"-instance of PARTITION. □

## 7.2 Annexation

Felsenthal and Machover (1998) prove that annexation is never disadvantageous with respect to the Shapley–Shubik index. For completeness, we give a simple proof of this fact.

**Proposition 18.** *For any weighted voting game $G$ with the set of players $N$, any $i \in N$, and any $S \subseteq N \setminus \{i\}$ we have $\varphi_i(G) \leq \varphi_{\&(S \cup \{i\})}(G_{\&(S \cup \{i\})})$.*

*Proof.* We give a proof for the case $|S| = 1$, i.e., $S = \{j\}$ for some $j \in N \setminus \{i\}$; the general case follows easily by induction. Let $\Pi_i$ be the set of all permutations $\pi$ of $N$ such that $i$ is critical for $\pi$ in $G$; we have $\varphi_i(G) = |\Pi_i|/n!$. For each $\pi \in \Pi_i$, let $f(\pi)$ be a permutation of the players in $N_{\&\{i,j\}}$ obtained from $\pi$ by deleting $j$ and replacing $i$ with a new player $\&\{i, j\}$. For any $\pi \in \Pi_i$ the player $\&\{i, j\}$ is pivotal for $f(\pi)$. Moreover, for any permutation $\sigma$ of $N_{\&\{i,j\}}$ we have $|f^{-1}(\sigma)| = n$. Hence, we have

$$\varphi_{\&\{i,j\}}(G_{\&\{i,j\}}) \geq \frac{|\{f(\pi) \mid \pi \in \Pi\}|}{(n-1)!} = \frac{|\Pi|/n}{(n-1)!} = \varphi_i(G).$$

□

However Felsenthal and Machover (1998) show that, for the case of the Banzhaf index, annexation could be disadvantageous; they refer to this phenomenon as the *Bloc Paradox*. They provide a 13-player WVG for which this is the case, which is the simplest example they could find. We improve on their result by describing a 7-player WVG where annexation is disadvantageous.

**Example 19.** Consider a weighted voting game $[11; 6, 5, 1, 1, 1, 1, 1]$. In this game, player 1 is pivotal for any coalition that involves player 2 and any subset of the remaining players, as well as for the coalition $\{3, \ldots, 7\}$, i.e., for 33 coalitions. Player 2 is pivotal for any coalition that involves player 1 and at most 4 of the remaining players, i.e., 31 coalitions. Finally, each of the players of weight 1 is only pivotal for the coalition that includes player 1 and the rest of the players of weight 1. Thus, the Banzhaf index of player 1 equals $\frac{33}{33+31+5} \approx 0.47826$.

If player 1 annexes one of the players with weight 1, the new game is $[11; 7, 5, 1, 1, 1, 1]$. Applying the same reasoning as above, we obtain that player 1 is pivotal for 17 coalitions, player 2 is pivotal for 15 coalition, and each of the remaining players is pivotal for exactly one coalition, so the Banzhaf index of player 1 in the new game is $\frac{17}{17+15+4} \approx 0.47222 < 0.47826$.

We have shown that annexation can be disadvantageous in the case of the Banzhaf index. One would at least expect that the Banzhaf index payoff after annexing another player is monotone in the power of the annexed player. Surprisingly, this is not the case. That is, we will now show that there exists a weighted voting game $G = [q; w_1, \ldots, w_n]$ and $i, j, k \in \{1, \ldots, n\}$ such that $w_j > w_k$, but $\beta_{\&\{i,j\}}(G_{\&\{i,j\}}) < \beta_{\&\{i,k\}}(G_{\&\{i,k\}})$. We will refer to this phenomenon as the *Annexation Non-monotonicity Paradox*. Observe that it is distinct from the Bloc Paradox: the former has to do with choosing which of the two given players to annex, while the latter has to do with choosing between annexing a given player and not annexing any player at all.

**Example 20.** Consider the weighted voting game $[9; 3, 3, 2, 1, 1, 1]$. Suppose first that player 1 annexes player 2 to form the game $[9; 6, 2, 1, 1, 1]$. In this game, player 1 is pivotal for 1 coalition that does not include player 2, and 7 coalitions that include player 2, i.e., 8 coalitions. Further, player 2 is pivotal for 6 coalitions, and each of the remaining players is pivotal for 2 coalitions. Thus, the Banzhaf index of player 1 is $\frac{8}{8+6+6} = 0.4$.

Now, suppose that player 1 annexes player 3 to form the game $[9; 5, 3, 1, 1, 1]$. In this game, player 1 is pivotal for 7 coalitions, player 2 is pivotal for 7 coalitions, and each of the remaining players is pivotal for 1 coalition. Thus, the Banzhaf index of player 1 in this game is $\frac{7}{7+7+3} \approx 0.411765 > 0.4$.

In contrast, the Shapley–Shubik index is monotone with respect to annexation.

**Proposition 21.** *For any weighted voting game* $G = [q; w_1, \ldots, w_n]$ *and any* $i, j, k \in \{1, \ldots, n\}$ *such that* $w_j \geq w_k$ *we have* $\varphi_{\&\{i,j\}}(G_{\&\{i,j\}}) \geq \varphi_{\&\{i,k\}}(G_{\&\{i,k\}})$.

*Proof.* Consider any permutation $\pi$ of $N_{\&\{i,k\}}$ for which $\&\{i, k\}$ is pivotal. Let $\pi'$ be a permutation of $N_{\&\{i,j\}}$ obtained by replacing $\&\{i, k\}$ with $\&\{i, j\}$ and $j$ with $k$. Since $w_{\&\{i,j\}} \geq w_{\&\{i,k\}}$, if in $\pi$ player $j$ appears after player $\&\{i, k\}$, then player $\&\{i, j\}$ is pivotal for $\pi'$. On the other hand, if in $\pi$ player $j$ appears before player $\&\{i, k\}$, we have $w(S_{\pi'}(\&\{i,j\})) \leq w(S_\pi(\&\{i,k\})) < q$, $w(S_{\pi'}(\&\{i,j\}) \cup \{\&\{i,j\}\}) = w(S_\pi(\&\{i,k\}) \cup \{\&\{i,k\}\}) \geq q$, so $\&\{i,j\}$ is pivotal for $\pi'$ in this case as well. Hence, each permutation for which $\&\{i,k\}$ is pivotal corresponds to a distinct permutation for which $\&\{i,j\}$ is pivotal, i.e., we have $\varphi_{\&\{i,j\}}(G_{\&\{i,j\}}) \geq \varphi_{\&\{i,k\}}(G_{\&\{i,k\}})$. □

To bound the gains and losses from annexation, we observe that a player can increase his payoff (with respect to both indices) by as much as 1. This happens if a dummy player annexes a sufficiently large player or a coalition of players. On the other hand, Theorem 7 immediately implies that, when a player $i$ annexes a player $j$ in a game $G$, we have $\beta_{\&\{i,j\}}(G_{\&\{i,j\}}) \geq \frac{1}{2}(\beta_i(G) + \beta_j(G))$. This has the following useful corollary.

**Corollary 22.** *For any weighted voting game $G$ with the set of players $N$ and any $i, j \in N$ we have $\beta_{\&\{i,j\}}(G_{\&\{i,j\}}) \geq \frac{1}{2}\beta_i(G)$, i.e., no player can decrease his payoff by more than a factor of 2 by annexing another player. Moreover, if $w_i \leq w_j$, then $\beta_{\&\{i,j\}}(G_{\&\{i,j\}}) \geq \beta_i(G)$.*

We will now show that determining whether a player can benefit from annexing a given coalition (with respect to the Banzhaf index) is NP-hard.

**Theorem 23.** BENEFICIAL-BI-ANNEXATION *is NP-hard.*

*Proof.* The proof is similar to that of Theorem 11. Given an instance of PARTITION $A = \{a_1, \ldots, a_k\}$, we construct a weighted voting game $G = [q; w_1, \ldots, w_n]$ with $n = k+2$ players as follows. We let $X = \sum_{a_i \in A} a_i$, and set $w_i = 8a_i$ for $i = 1, \ldots, n-2$, $w_{n-1} = w_n = 1$ and $q = 4X + 2$.

By Lemma 10, if $A$ is a "no"-instance of PARTITION, then players $n-1$ and $n$ are dummies, and $n$ remains a dummy even if it annexes $n-1$. Now, suppose that $A$ is a "yes"-instance of PARTITION. Let $x$ denote the number of coalitions in $N \setminus \{n-1, n\}$ of weight $4X$. We have $\eta_{n-1} = \eta_n(G) = x$. For $i = 1, \ldots, n-2$, let

$$\mathcal{S}_i = \{S \subseteq N \setminus \{n-1, n, i\} \mid w(S) < 4X, w(S) + w_i \geq q\},$$

and set $y_i = |\mathcal{S}_i|$. Also, set $y = \sum_{i=1}^{n-2} y_i$.

The calculations in the proof of Theorem 11 show that

$$\beta_n(G) = \frac{x}{2x + (n-2)\frac{x}{2} + 4y},$$
$$\beta_{\&\{n,n-1\}}(G_{\&\{n,n-1\}}) = \frac{x}{x + (n-2)\frac{x}{2} + 2y}.$$

Since $x > 0$, this implies $\beta_n(G) < \beta_{\&\{n,n-1\}}(G_{\&\{n,n-1\}})$. Hence, $(G, \{n-1\}, n)$ is a "yes"-instance of BENEFICIAL-BI-ANNEXATION if and only if we have started with a "yes"-instance of PARTITION. □

We conclude this section by analyzing the benefits of merging and annexation in unanimity games, i.e., games where $q = w(N)$.

**Proposition 24.** *In any unanimity game, it is advantageous for a player to annex an arbitrary coalition, with respect to both the Shapley–Shubik index and the Banzhaf index. However, no group of players can increase its total payoff (as measured by either of the indices) by merging.*

*Proof.* In any game all players have equal value of the index both before and after annexation. Hence, if $i$ annexes a coalition of size $s$, his power increases from $\frac{1}{n}$ to $\frac{1}{n-s}$. However, merging reduces the total power of all players in a coalition of size $s$ from $\frac{s}{n}$ to $\frac{1}{n-s}$. $\qquad\square$

We remark that Proposition 24 generalizes to any game in which each player is a veto player.

## 8. Empirical Analysis

We now analyze false-name splitting manipulations empirically. We have constructed a system for randomly constructing weighted voting games and examined the changes in both the Shapley–Shubik index and the Banzhaf index that occur when agents split their weights between false identities. We briefly describe our simulation system, the game construction and the power index calculations, and then present the empirical evidence obtained.

### 8.1 Simulation System and Settings

The weighted voting games were constructed by first randomly choosing the number of players in the game. Then, the weights of the players were each drawn from $N(\mu, \sigma^2)$, the normal distribution with mean $\mu$ and variance $\sigma^2$. The weights were then rounded to the nearest integer, to make sure that the game has integer weights. The threshold for the game was chosen uniformly at random between 0 and the sum of the players' weights $w = w(N)$, and again rounded to the nearest integer. In our experiments, we have used a mean of $\mu = 200$ for the weights, and several values for the standard deviation $\sigma$ from the set $\{5, 10, 15, \ldots, 50\}$. The number of players $n$ was chosen uniformly at random from the set $\{5, 6, 7, \ldots, 24\}$.

Power indices are computationally hard to compute exactly (Papadimitriou & Yannakakis, 1994; Matsui & Matsui, 2001), but can be tractably approximated using several methods. We have used the approximation method of Bachrach et al. (2010). This algorithm estimates the power indices and returns a result which is probably approximately correct, as discussed in Section 6. Given a game in which a player's true power index is $\psi$, and given a target accuracy level $\epsilon$ and confidence level $\delta$, the algorithm returns an approximation $\hat{\psi}$ such that with probability at least $1 - \delta$ we have $|\psi - \hat{\psi}| \leq \epsilon$ (i.e. the result is approximately correct, and is within a distance $\epsilon$ of the correct value). This algorithm works by drawing a sample of $k$ permutations (or coalitions), and testing whether the target player is critical for them. Such a test runs in time linear in the number of agents, so the total running time is $O(kn \log W)$. Bachrach et al. show that to achieve a confidence level $\delta$ and accuracy level $\epsilon$, it suffices to take $k = \ln(2/\delta)/(2\epsilon^2)$. Thus the total running time is logarithmic in the confidence and quadratic in the accuracy, so the approach is tractable even for high accuracy and confidence. We have used $\delta = 0.00001$ and $\epsilon = 0.001$, so the power was estimated very accurately. Our system was implemented in C#, and the results of the experiments were stored in an SQL database. Since power indices were approximated accurately, a single experiment can take several seconds. Our tests required tens of thousands of experiments, so we have used a compute cluster with 250 cores for our experiments.

Our theoretical results show that testing for a beneficial split is hard, which might create the impression that finding a beneficial manipulation is hard in practice. Our empirical experiments were designed to see whether this is indeed the case. A very naive method that a manipulator can use is to try many possible splits into two identities, in constant intervals. In other words a manipulator whose weight is $w$ can try $\frac{s}{2}$ splits by splitting his weight as $(\frac{w}{s}, w - \frac{w}{s})$, $(\frac{2w}{s}, w - \frac{2w}{s})$, $(\frac{3w}{s}, w - \frac{3w}{s})$ and so on. Although this is certainly not a complete coverage of the space of possible manipulations, in our experiments we have tried a very simple algorithm that is based on this idea. Since all the weights were integers, we have only tried splitting weights between two false identities, and examined all integer splits. For example, if an agent had a weight of $w_i = 10$, we attempted splitting into weights of $w_1' = 9, w_1'' = 1$, $w_1' = 8, w_1'' = 2$, $w_1' = 7, w_1'' = 3$ and so on. In each experiment we have recorded the details of the game, the number of beneficial splits (power increase) and harmful splits (power decrease). For a split to be considered beneficial, it had to increase the power by more than twice the accuracy level. Thus, the results presented here understate the number of positive splits. In our results we examine the proportion of experiments where we have found at least one beneficial manipulation, as well as the proportion of the splits that were beneficial (out of all the integer splits).

## 8.2 Empirical Results

We first present the results regarding the Shapley–Shubik index. First and foremost, our results indicate that the weighted voting domain is very manipulable, at least for our method of generating random weighted voting games. Under all values we have tried for the variance in the player weights and number of players in the game, in over $95.5\%$ of the experiments even the very naive manipulation algorithm managed to uncover at least one beneficial manipulation. This indicates that in most games it is enough to try all integer splits (or splits in uniform intervals) and use the tractable method for approximating power indices to uncover beneficial manipulations. Figures 2 and 3 indicate the proportion of the experiments in which this algorithm succeeds in finding a beneficial split, as a function of the variance in players' weights and the number of players, respectively. It appears that the success rate of our algorithm slightly increases as the variance increases. No such obvious trend appears for the number of players.

One might be tempted to think that beneficial splits are quite common, as most experiments had at least one beneficial split. However, it turns out that most splits are harmful splits. In all tested settings, less than $40\%$ of all splits were beneficial splits. In most settings, harmful splits accounted for over $70\%$ of all splits. In Figure 4 and Figure 5, we indicate the proportion of beneficial splits, as a function of the variance in players' weights and the number of players, respectively.

We have also examined the distribution of the proportion of beneficial splits across experiments. In some generated games, the beneficial splits were quite rare, and less than a single percent of the splits were beneficial. In other generated games, beneficial splits were the common case, and over $99\%$ of the splits were beneficial. Figure 6 shows the distribution (histogram) of the proportion of beneficial splits, across games. To create the figure, the games were partitioned into 200 bins, according to the proportion of beneficial splits in the game. Each bin was of size $0.5\%$ (e.g., the proportion of beneficial manipulations in the 100-th bin was between 0.5 and 0.55). The value on the X axis of Figure 6 is the proportion of beneficial splits (the bin), and the Y axis is the number of experiments falling in that category. Figure 6 shows that most games are ones where beneficial splits are more rare than harmful splits, but the distribution has a long tail, so even games where almost all the splits are beneficial are not uncommon.

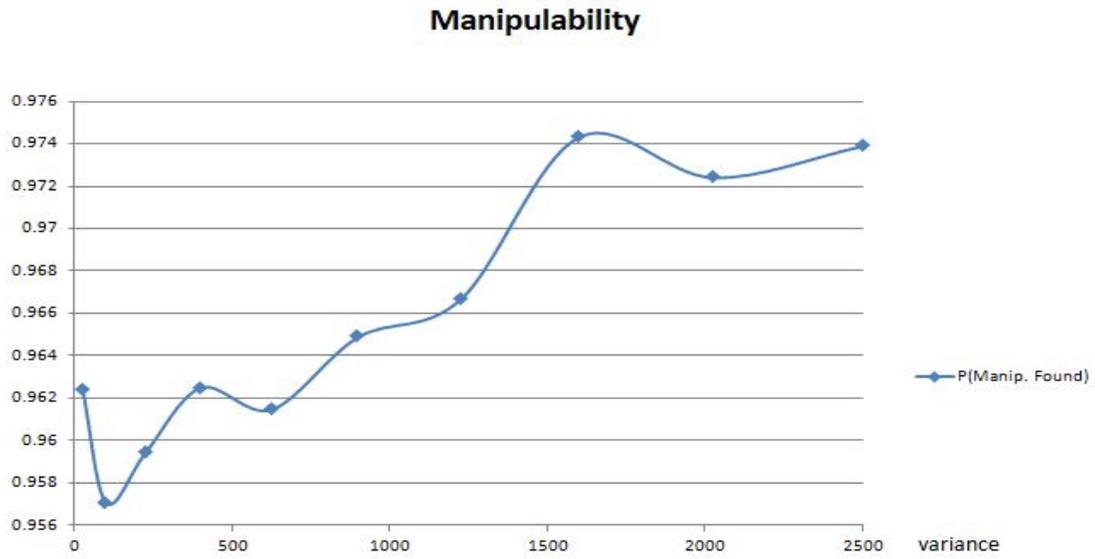

Figure 2: Proportion of experiments where the naive algorithm finds a beneficial split for different variances of players' weights (Shapley–Shubik index)

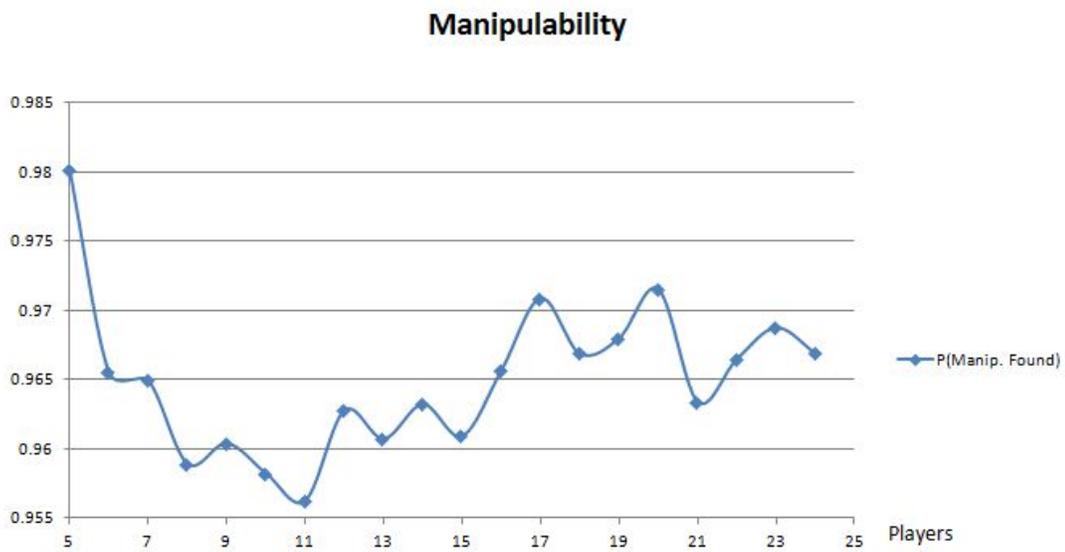

Figure 3: Proportion of experiments where the naive algorithm finds a beneficial split for different numbers of players (Shapley–Shubik index)

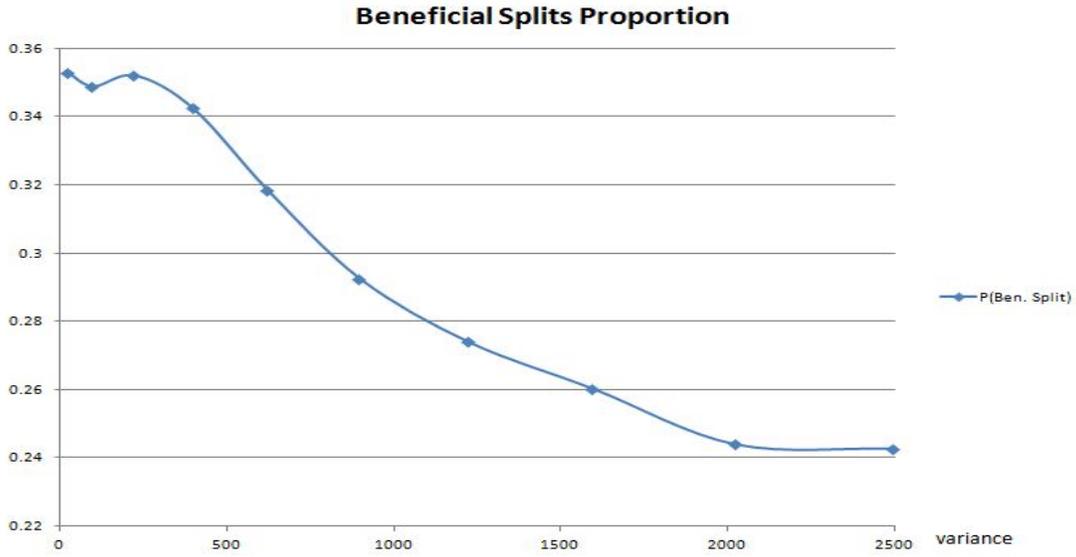

Figure 4: Proportion of beneficial splits for different variances of players' weights (Shapley–Shubik index)

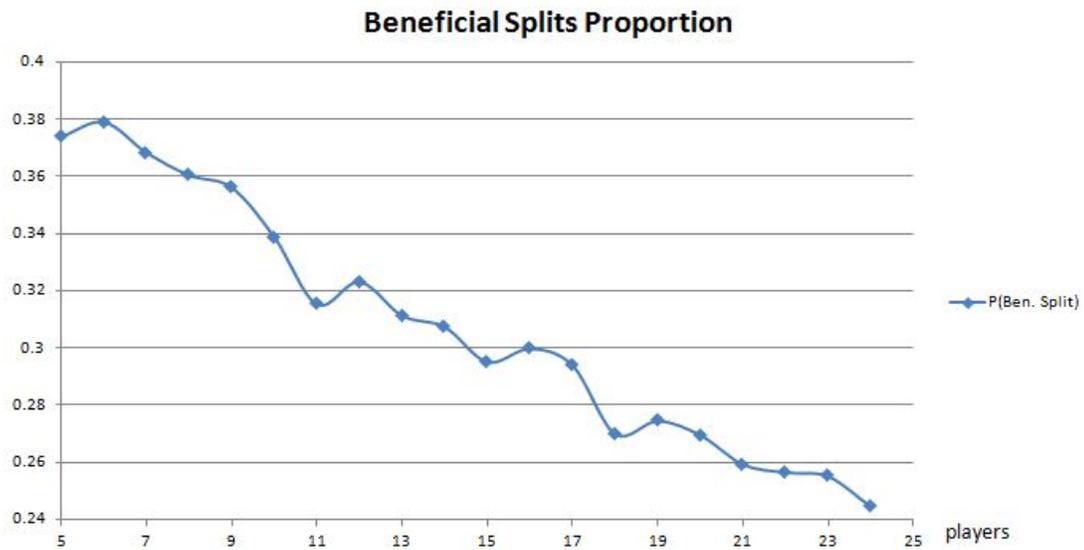

Figure 5: Proportion of beneficial splits for different numbers of players (Shapley–Shubik index)

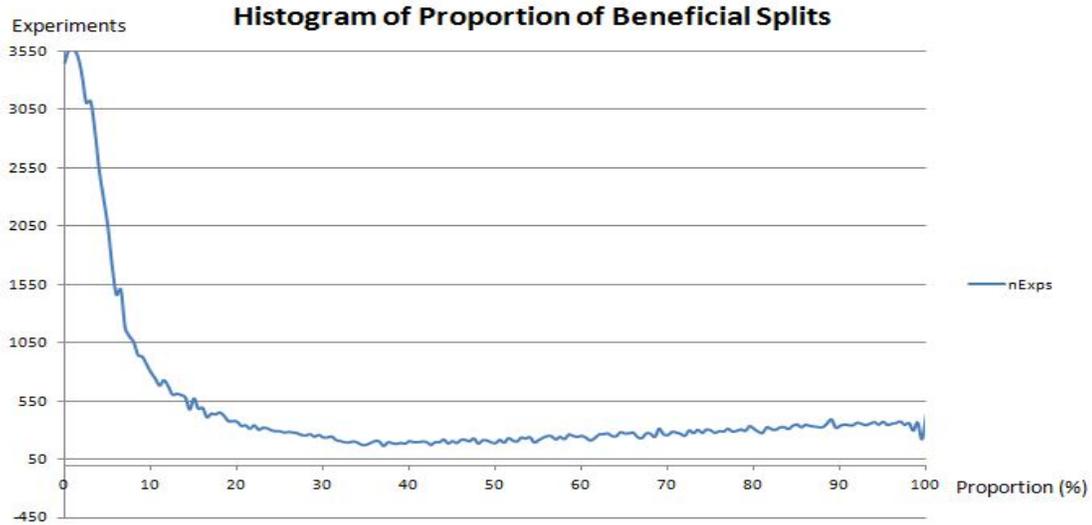

Figure 6: Distribution (histogram) of the proportion of positive splits across games (Shapley–Shubik index)

We now turn to examine the Banzhaf index. As in the case of the Shapley–Shubik index, for the Banzhaf index the weighted voting domain is very susceptible to manipulation. Under all tested settings, in over 92.5% of the experiments our manipulation algorithm managed to uncover at least one beneficial manipulation (slightly less than the 95.5% for the Shapley–Shubik index). The proportion of experiments where our algorithm finds a beneficial split with respect to the Banzhaf index is shown in Figure 7 (for different values of variance) and Figure 8 (for different number of players).

Similarly to the case of the Shapley–Shubik index, for the Banzhaf index beneficial splits were the less common case, and most splits are harmful splits, with less than 45% of all splits being beneficial, and typically about 40% being beneficial splits (slightly higher than for the Shapley–Shubik index). Unlike for the Shapley–Shubik index, for the Banzhaf index, the proportion of beneficial splits among all splits increases with the variance (Figure 9). However, this proportion does not have a clear trend with regard to the number of players (Figure 10).

The distribution of the proportion of beneficial splits across games for the Banzhaf index seems quite similar to that for the Shapley–Shubik index (see Figure 11). Again, for most games, the majority of splits are harmful, but the distribution has a long tail, and many games have mostly beneficial splits. Although the distribution seems similar to that for the Shapley–Shubik index, the tail of the distribution seems slightly "fatter" for the Banzhaf index.

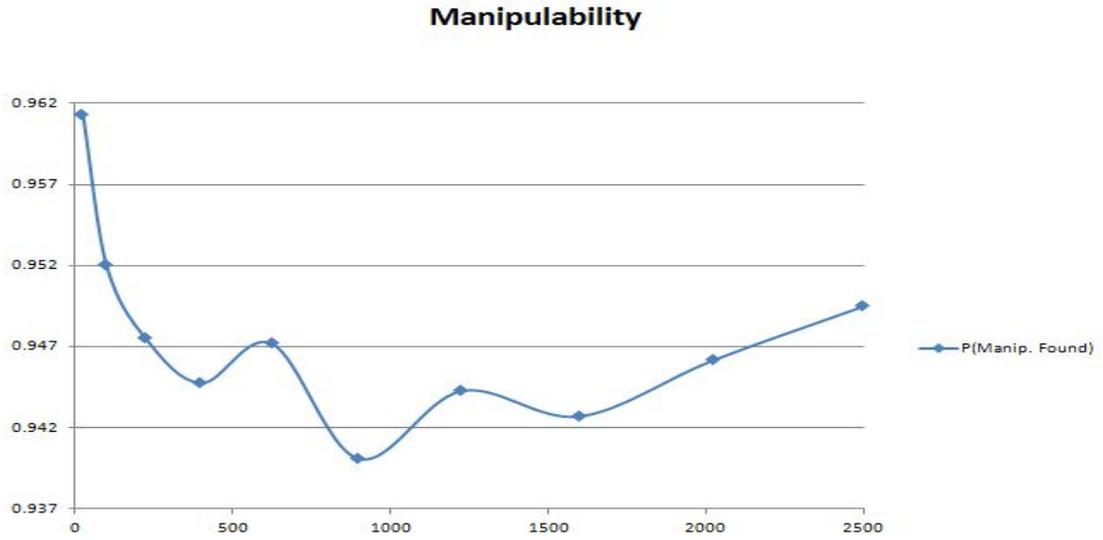

Figure 7: Proportion of experiments where the naive algorithm finds a beneficial split for different variances of players' weights (Banzhaf index)

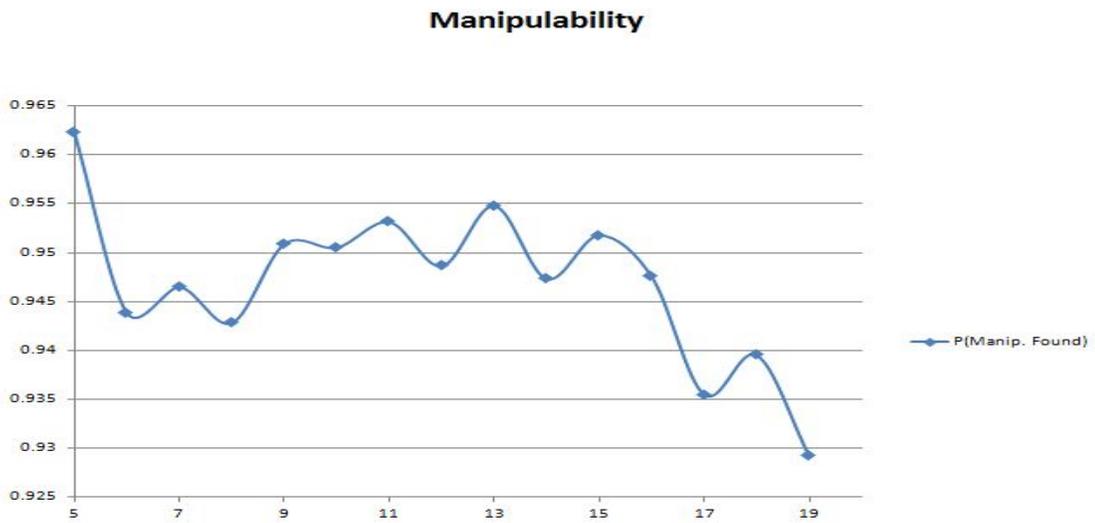

Figure 8: Proportion of experiments where the naive algorithm finds a beneficial split for different numbers of players (Banzhaf index)

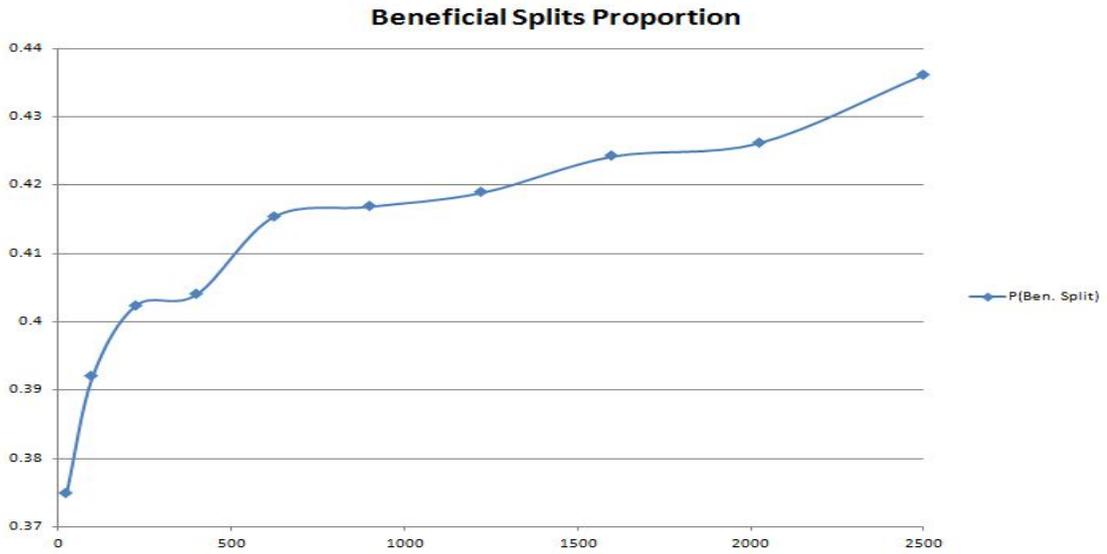

Figure 9: Proportion of beneficial splits for different variances of players' weights (Banzhaf index)

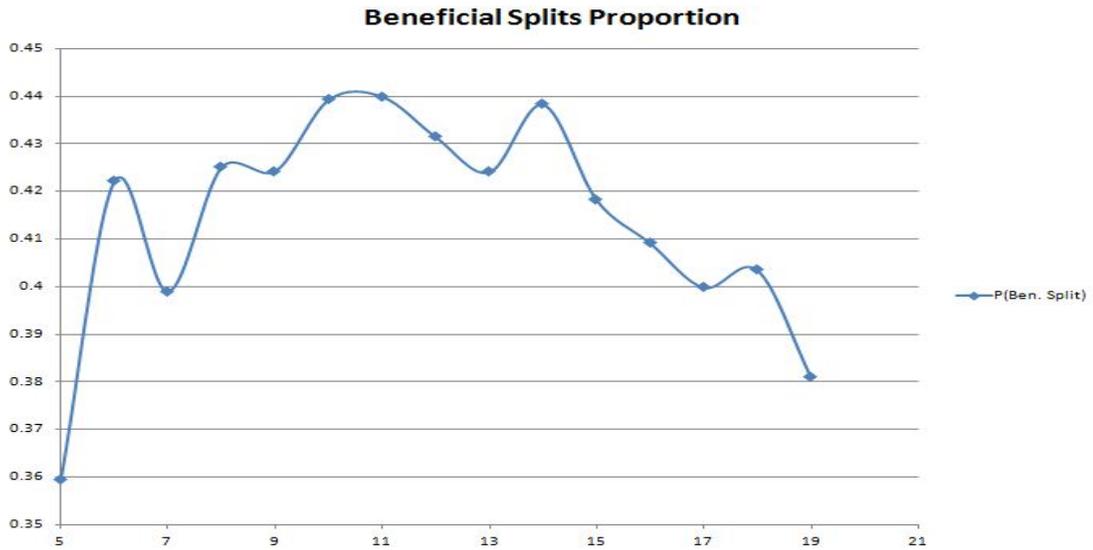

Figure 10: Proportion of beneficial splits for different numbers of players (Banzhaf index)

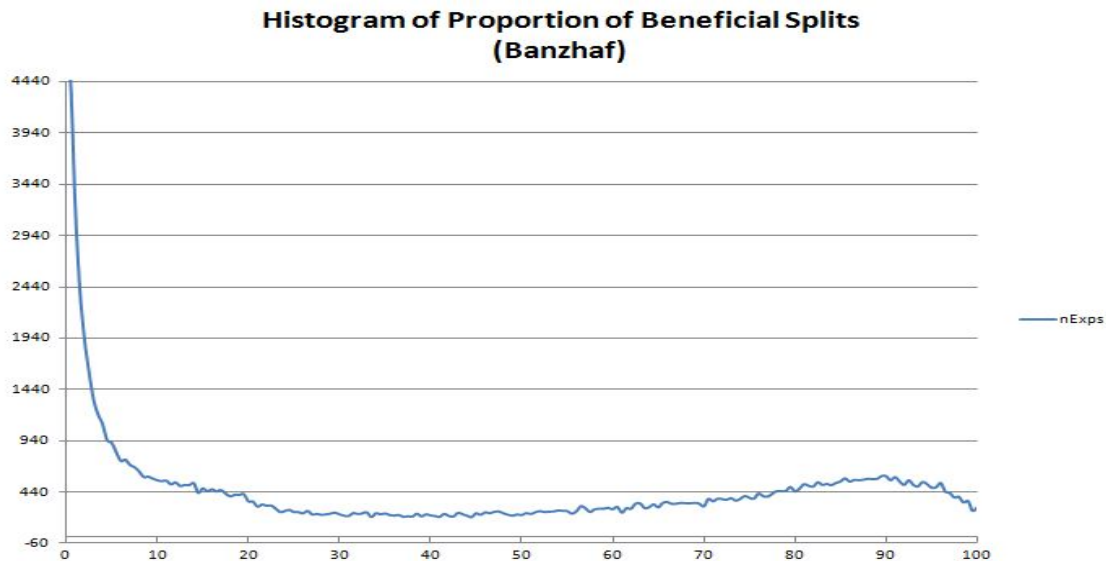

Figure 11: Distribution (histogram) of the proportion of positive splits across games (Banzhaf index)

To conclude, experiments for both indices present a similar picture. For most games generated in our model, we have mostly harmful splits. However, many experiments have many positive splits, and in some of them even almost all the splits are beneficial. Games where trying all the integer splits does not yield a successful manipulation are very rare, although they do exist. Thus, for most games generated in our model, even the extremely simple manipulation algorithm finds beneficial splits. We conclude that despite the hardness results in this paper, in practice we believe it is quite easy to find such splits, and thus we believe such attacks pose a real problem in many settings.

## 9. Splitting into More Than Two Identities

So far, we have mostly discussed the gain (or loss) that a player can achieve by splitting into two identities. However, it is also possible for a player to use three or more false names. Potentially, the number of identities a player can use can be as large as his weight (and if the weights are not required to be integers, it can even be infinite). It would be interesting to see which of our results hold in this more general setting. For example, while our computational hardness result holds for splits into any number of identities, the algorithmic results of the previous section only apply to splits into a constant number of new identities. An obvious open problem here is to design a pseudopolynomial algorithm for finding a beneficial integer split into any number of identities, or to prove that this problem is NP-hard even for small weights (i.e., weights that are polynomial in $n$). Another question of interest here is to extend the upper and lower bounds of Section 4 for this setting.

One might think that finding a beneficial split into $k \geq 2$ identities is easier than finding one that uses exactly two identities: after all, any two-way split can be transformed into a split into three or more players in which only two players have non-zero weight. However, it turns out that if we restrict our attention to non-trivial splits, i.e., one in which all of the new players have a non-zero weight, this is no longer the case.

**Example 25.** Consider a game $G = [6; 5, 5]$. In this game, the only winning coalition includes both players, so their Shapley–Shubik indices are given by $\varphi_1(G) = \varphi_2(G) = 1/2$. Suppose that player 2 splits into two identities $2'$ and $2''$. For any selection of integer weights $w_2' > 0, w_2'' > 0$ that satisfy $w_2' + w_2'' = 5$, in the new game $G' = [6; 5, w_2', w_2'']$ we have $\varphi_{2'}(G') = \varphi_{2''}(G') = 1/3$. Indeed, in this game each player is pivotal in a permutation $\pi$ if and only if it occurs in the second position, which happens with probability $1/3$. Hence, any non-trivial split into two identities increases the payoff of the second player by a factor of $(2/3)/(1/2) = 4/3$.

Now, suppose that the second player splits into 5 new players of weight 1 each. In the new game, player 1 is pivotal for a permutation $\pi$ if and only if he does not occur in the first position in that permutation, so his Shapley–Shubik index is $5/6$. Consequently, the sum of Shapley–Shubik indices of all remaining players (i.e., the new identities of player 2) is $1/6$. Therefore, this split decreases the payoff of player 2 by a factor of 3. To summarize, while any non-trivial integer split into 2 identities is beneficial for player 2, any integer split into 5 identities with positive weight is harmful for him.

**Remark 26.** Example 25 can be generalized to games of the form $G_N = [N + 1; N, N]$ for an arbitrary integer $N > 0$. The reasoning above shows that if one of the players decides to split into $N$ new players of weight 1 each, this increases the Shapley–Shubik index of the other player to $N/(N + 1)$ and hence decreases the total payoff of the splitting player by a factor of $(N + 1)/2$. As the representation size of this game is polynomial in $\log N$, this decrease is *exponential* in description size.

## 10. Conclusions

We have considered false-name manipulations in weighted voting games with respect to the payoff schemes based on the Shapley–Shubik index and the Banzhaf index. We have also considered manipulation via annexation and voluntary merging with respect to such payoff schemes. We have examined both the limits of manipulation (Table 1) and the complexity of manipulation (Table 2), and complemented the theoretical investigation by empirical analysis.

We have shown that, in most scenarios considered in this paper, testing whether a beneficial manipulation exists is NP-hard. One may ask whether these hardness results provide an adequate barrier to manipulation, given that the power indices themselves are hard to compute. In other words, don't we simultaneously assume that the weights are small (and hence computing the indices is easy) and large (and hence manipulation is hard)? To resolve this apparent contradiction, note that the power indices considered in this paper

| Bounds | Reference |
|---|---|
| $\frac{2}{n+1}\varphi_i(G) \leq \varphi_{i'}(G') + \varphi_{i''}(G') \leq \frac{2n}{n+1}\varphi_i(G)$ | Theorems 5 and 6 |
| $\varphi_i(G) \leq \varphi_i(G_{\&(\{i\}\cup S)}) \leq 1.$ | Proposition 18 |
| $\frac{1}{n+1}\beta_i(G) \leq \beta_{i'}(G') + \beta_{i''}(G') \leq 2\beta_i(G)$ | Theorems 7 and 8 |
| $\frac{\beta_i(G)}{2} \leq \beta_i(G_{\&(\{i,j\})}) \leq 1.$ | Corollary 22 |

Table 1: Bounds on effects of false-name manipulations in WVGs

| | Banzhaf index | Shapley–Shubik index |
|---|---|---|
| SPLITTING | NP-hard | NP-hard |
| MERGING | NP-hard | NP-hard |
| ANNEXATION | NP-hard | advantageous* |
| SPLITTING in unanimity game | advantageous | advantageous |
| MERGING in unanimity game | disadvantageous | disadvantageous |
| ANNEXATION in unanimity game | advantageous | advantageous |

*(Felsenthal & Machover, 1998)

Table 2: Complexity of false-name manipulations in WVGs

correspond to the voting power, and the players may try to increase their voting power by weight-splitting manipulation even if they cannot compute it. Also, when a power index is used to compute payments, the center, which performs this computation, may have more computational power than individual players.

Our experimental results show that, for moderately large weights, weight-splitting manipulation is easy in practice. However, our algorithm relies on considering all integer splits, i.e., its running time is at least linear in the manipulator's weight. An interesting open question is whether it is the case that if a beneficial split exists, it can be found by testing a number of splits that is logarithmic in the manipulator's weight.

Our results indicate that the Shapley–Shubik index and the Banzhaf index behave similarly with respect to false-name manipulation; however, the Shapley–Shubik index appears to be a more desirable solution concept because annexation does not decrease the payoff of a player. Exploring other solution concepts and their behavior with respect to false-name manipulation is a natural next step; a particularly suitable solution to consider could be the nucleolus, which not only always exists but is also unique.

The study of weighted voting has many applications, both in political science and in multiagent systems. There are several possible interpretations for identity-splitting in these contexts, such as obtaining a higher share of the grand coalition's gains when these are distributed according to the Shapley–Shubik index or the Banzhaf index, or obtaining more political power by splitting a political party into several parties with similar political platforms. In the first case, a false-name manipulation is hard to detect in open anonymous environments, and can thus be very effective. In the second case, the manipulation is done using legitimate tools of political conduct. Therefore, we conjecture that false-name manipulation is widespread in the real world and may become a serious issue in multiagent systems. It is therefore important to develop a better understanding of the effects of this behavior and/or design methods of preventing it.

## Acknowledgments

Haris Aziz and Mike Paterson were partially supported by DIMAP (the Centre for Discrete Mathematics and its Applications). DIMAP is funded by the UK EPSRC under grant EP/D063191/1. Partial support for Aziz's research was also provided by the Deutsche Forschungsgemeinschaft under grants BR-2312/6-1 (within the European Science Foundation's EUROCORES program LogICCC) and BR 2312/3-2. Edith Elkind was partially supported by ESRC under grant ES/F035845/1 and by Singapore NRF Research Fellowship 2009-08.